\documentclass[]{spie}  

\usepackage{amsmath,amsfonts,amssymb}
\usepackage{graphicx}
\usepackage[colorlinks=true, allcolors=blue]{hyperref}
\usepackage{booktabs}

\title{CCAT: Design and Characterization of the 350 GHz Instrument Module}

\author[a]{Ben Keller}
\author[b]{Jordan Wheeler}
\author[a]{Cody J. Duell}
\author[a]{Darshan A. Patel}

\author[b]{Jason Austermann}
\author[c]{Baird D. Bankovic}
\author[d]{James Burgoyne}
\author[d,e]{Scott Chapman}
\author[f]{Steve K. Choi}
\author[c]{Rodrigo Freundt}
\author[a]{Min Gao}
\author[f]{Eliza Gazda}
\author[d]{Anthony I. Huber}
\author[b]{Johannes Hubmayr}
\author[a]{Lawrence T. Lin}
\author[f]{Quintin Meyers}
\author[f]{Paul Malachuk}
\author[a]{Alicia Middleton}
\author[a,c]{Michael D. Niemack}
\author[g]{Tilak M. Patel}
\author[b,h]{Anna Vaskuri}
\author[a,g]{Eve Vavagiakis}
\author[b]{Michael R. Vissers}
\author[a]{Samantha Walker}
\author[a]{Yuhan Wang}
\author[d]{Ruixuan (Matt) Xie}

\affil[a]{Department of Physics, Cornell University, Ithaca, NY 14850, USA}

\affil[b]{National Institute of Standards and Technology, Quantum Sensors Division, Boulder, CO 80305, USA}

\affil[c]{Department of Astronomy, Cornell University, Ithaca, NY 14850, USA}

\affil[d]{Department  of Physics and Astronomy, University of British Columbia, Vancouver, Canada}

\affil[e]{Department of Physics and Atmospheric Science, Dalhousie University, Halifax, BC, Canada
}
\affil[f]{Center for Experimental Cosmology and Instrumentation, Department of Physics and Astronomy, University of California, Riverside, CA 92521, USA}

\affil[g]{Department of Physics, Duke University, Durham, NC 27704, USA}

\affil[h]{Department of Physics, University of Colorado, Boulder, Colorado, United States Colorado, United States}

\authorinfo{Further author information: (Send correspondence to B.K.)\\B.K.: E-mail: bdk54@cornell.edu}

\begin{document} 
\maketitle

\begin{abstract}
The CCAT Collaboration’s Prime-Cam instrument will soon be deployed to the Fred Young Submillimeter Telescope (FYST) in Chile’s Atacama Desert. Featuring prominently in Prime-Cam’s calibration and early science observations will be the 350 GHz instrument module, a broadband camera that will field more than 10,000 microwave kinetic inductance detectors (KIDs) across three detector arrays. Forecasts show this module will be capable of making the most sensitive to-date measurements of polarized dust emission over a large fraction of the sky at this frequency, enabling new galactic polarization science and improved understanding of cosmological foregrounds \cite{ccatsciencepaper}. In this work we discuss the design of the 350 GHz instrument module, covering aspects of the optics, readout, and detector arrays. We then report on the results of in-lab testing of the fully-integrated module, achieving stable cryogenic performance with a 100~mK focal plane, high detector yield, and a passband comparable to designed specifications. Upon completion of these tests, this module was shipped to the telescope site in Chile for integration in Prime-Cam.
\end{abstract}


\section{INTRODUCTION}
\label{sec:intro} 

    \begin{figure}
    \begin{center}
    \includegraphics[width=0.5\linewidth]{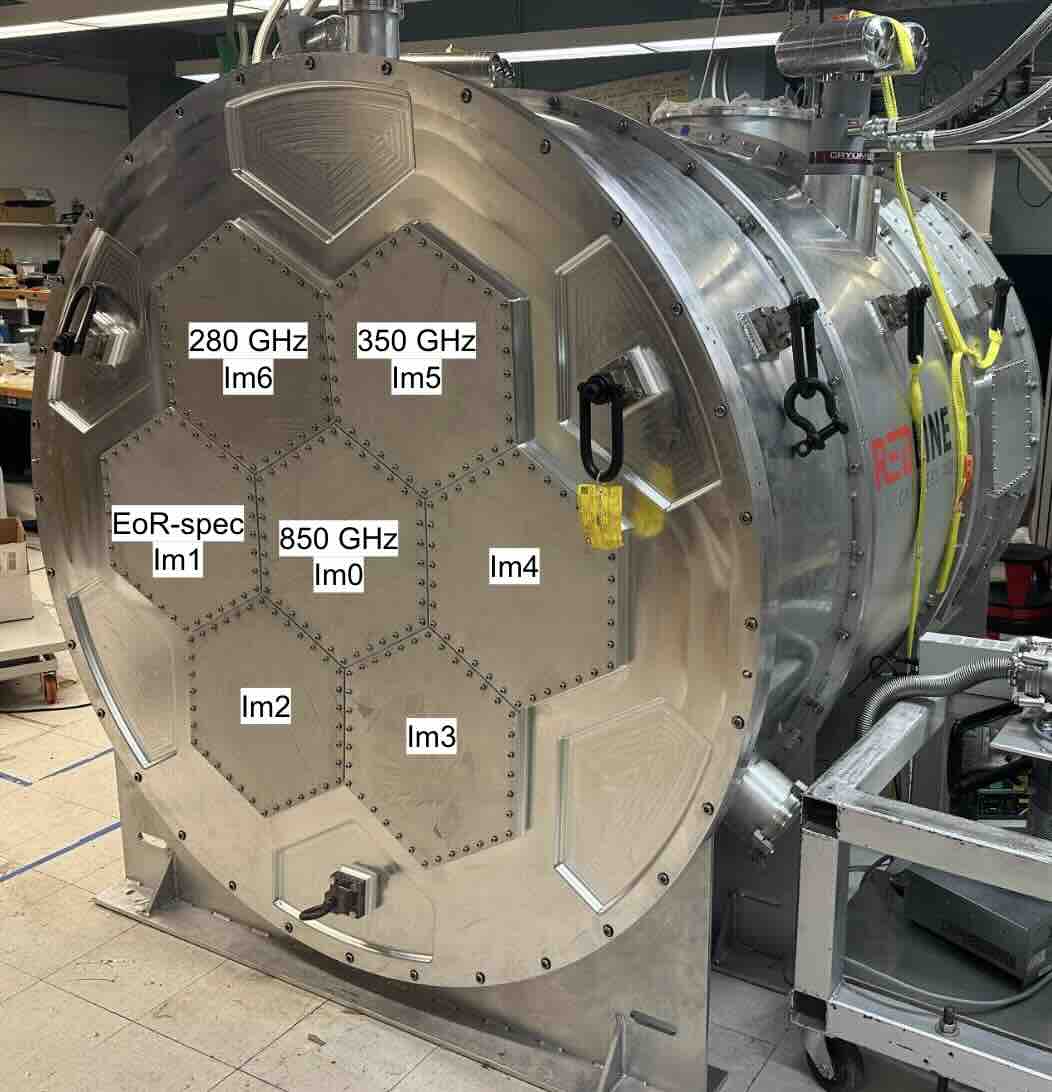}
    \end{center}
    \label{fig:module_locations} 
    \caption{Planned positions for the first 4 instrument modules in Prime-Cam. Initial observations will proceed with the 280~GHz and 350~GHZ broadband modules. The 850~GHz broadband and Epoch of Reionization Spectrometer (EoR-spec) module will be deployed after validation and testing in Mod-Cam. Development of future module, including a 410~GHz broadband module, is also underway \cite{tilak_450_inprep}. Module locations were chosen based on a number of mechanical considerations as well as optical simulations \cite{Huber:2025elf}.}
   \end{figure} 

The CCAT Observatory's Fred Young Submillimeter Telescope (FYST) is nearly assembled and is soon to undergo commissioning on Cerro Chajnantor in the Chilean Atacama Desert. FYST is a 6-meter aperture crossed-Dragone telescope located at an elevation of 5600 m, featuring a large 8 degree field-of-view at 3~mm and high surface accuracy (\textless 10.7 µm HWFE) mirrors \cite{Parshley_2018}. In the coming months the CCAT Observatory's primary science instrument, Prime-Cam, will be deployed to FYST. While Prime-Cam has the ability to field more than 100,000 kinetic inductance detectors (KIDs) in total across seven cameras, or ``instrument modules", initial observations will employ only two broadband polarimetric modules with observing passbands centered at 280~GHz and 350~GHz. Each of these modules will field approximately 10,000 KIDs across three detector arrays \cite{vavagiakis_modcam_2022, duell_280ghz}. 

When fully populated with instrument modules, Prime-Cam will target a wide range of science goals facilitated by its ability to make broadband polarimetric and spectroscopic measurements between 210--850~GHz \cite{vavagiakis_modcam_2022, rodrigo_eor, tilak_450_inprep, chapman_850} at a high-altitude site with exceptionally low atmospheric precipitable water vapor (PWV) \cite{pwv_study}. In particular, the 350~GHz module discussed here is forecasted to improve on the signal-to-noise ratio of current state-of-the-art Planck 353 GHz polarized dust foreground measurements by a factor of \textgreater~2 \cite{ccatsciencepaper}. These measurements, along with 410 and 850~GHz broadband observations coming online in subsequent observing seasons, will help to reduce bias on the tensor-to-scalar ratio \textit{r} and may enable experiments such as Simons Observatory (SO) to detect \textit{r} as low as $0.002$ \cite{ccatsciencepaper}.

To characterize the instrument modules prior to their deployment to FYST, we utilize a local single-module testbed receiver, Mod-Cam, which is capable of performing end-to-end testing in a variety of dark and optical configurations in a cryogenic environment similar to that of Prime-Cam \cite{vavagiakis_modcam_2022}. While extensive in-lab testing has been performed on the initial 280~GHz module \cite{patel2025ccatreadout, lin2025ccatmodcamcryogenicperformance}, this is the first work to report on the status of the 350~GHz module and detector arrays. In this work, we first outline the design of the 350 GHz instrument module components and detector arrays in Section \ref{sec:design} and then discuss the pre-deployment validation testing performed in Mod-Cam at Cornell University in Section \ref{sec:testing}.

\section{350 GHz Instrument Module Design}\label{sec:design}

   \begin{figure}
   \begin{center}
   \includegraphics[width=0.9\linewidth]{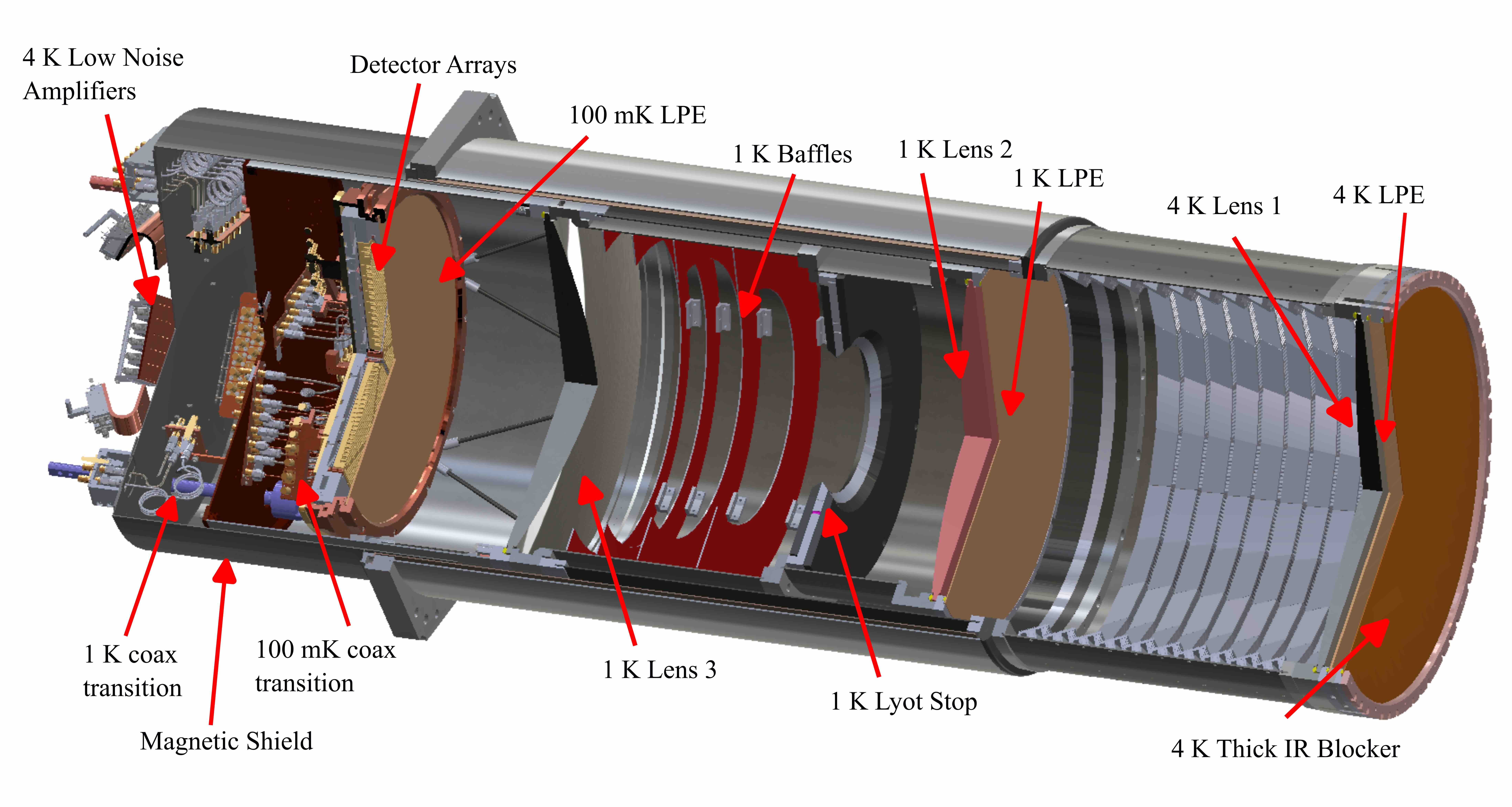}
   \end{center}
    \label{fig:350_module_cutaway}
   \caption{A 3D-model cutaway view of the 350~GHz instrument module design. Important optical elements such as lenses and filters, along with the detectors and readout are labeled.}
   \end{figure}

The 350 GHz instrument module mechanical design closely follows the 280 GHz module design and is based on the optics tube designs for the Simons Observatory Large Aperture Telescope Receiver \cite{vavagiakis_modcam_2022, Sierra_2025_SO_optics_tube}. The 350~GHz module, like all CCAT instrument modules, is mechanically mounted and cryogenically linked to the Prime-Cam 4~K stage. The planned early configuration of instrument modules within Prime-Cam is shown in Fig. \ref{fig:module_locations}, where module locations have been chosen based on optical simulations and considerations for the assembly and installation process of modules \cite{Huber:2025elf}. Each module acts as a nearly fully-integrated camera unit, with cryogenic stages at temperatures of 4~K, 1~K, and 100~mK housing optical elements, detectors, and readout. A 3D model of all components within the 350~GHz instrument module is shown in Fig. \ref{fig:350_module_cutaway}. 

\subsection{Instrument Module Optics}
\label{sec:optics}
Each instrument module has a series of sub-4~K optical elements mounted inside the module, along with a corresponding set of elements outside the module mounted to the Prime-Cam 300~K, 80~K, and 40~K shells. At the front of Prime-Cam, light first enters the receiver through a Ultra-high-molecular-weight polyethylene (UHMWPE) window with an anti-reflective (AR) coating and passes through a double sided metal mesh infrared blocking filter (DSIR Blocker) mounted at 300~K: a low-pass filter with a cutoff in the THz range. Light then passes through three additional filters mounted at 80~K: an alumina wedge filter sandwiched by two DSIR blockers. The 80~K alumina wedge is designed to reorient the telescope optics along the axis of the module and features a metamaterial AR coating ablated into each side. Light passes through one final 40~K DSIR before reaching the first of the 4~K module optics.

The front of the instrument module 4~K stage features a single thick IR-blocking filter. Three monolithic silicon lenses  (one at 4~K, two at 1~K) fabricated at the University of Chicago featuring ablated AR coatings are used to form a virtual image of light from the sky at the position of the detector arrays on the focal plane. At 4~K, 1~K, and 100~mK, a series of low-pass filters with successively decreasing cutoffs between 500~GHz--370~GHz are also installed. The 100~mK low-pass filter has the lowest frequency cutoff and sets the high edge of the module passband. This filter is therefore an especially critical optical element given the presence of an atmospheric water line immediately above the designed passband \cite{ALMA_handbook}.  

The instrument module additionally features elements designed to reduce stray light and eliminate possible diffraction effects at the focal plane. Inside the front 4~K tube, metamaterial injection-molded tiles originally designed for lower frequency Simons Observatory optics tubes have been installed \cite{Xu_metamaterial}. The design and physical properties of these tiles indicate they should maintain adequate absorption properties for the 350~GHz band; future higher-frequency instrument modules such as the 850~GHz module may consider alternative blackening methods. Inside the 1~K tube at the module pupil, a Lyot stop coated in similar tiles has been installed. A series of baffles blackened with Stycast 2850FT expands out behind the stop to further absorb stray light outside the geometric optical path. 

\subsection{Instrument Module Readout}
\label{sec:readout}

\begin{figure}
   \begin{center}
   \includegraphics[width=1\linewidth]{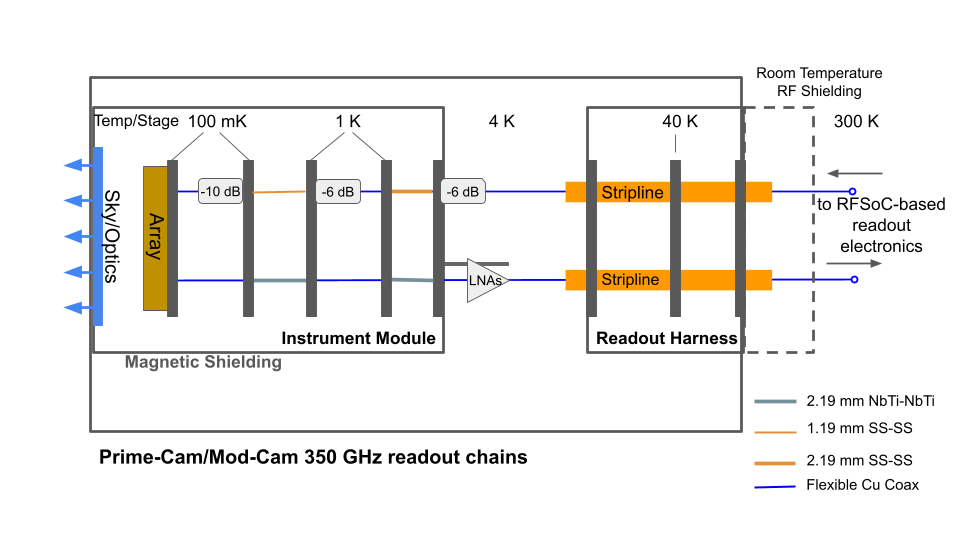}
   \end{center}
   \caption
   {A diagram of a single complete readout chain in the 350 GHz instrument module installed in either Prime-Cam or Mod-Cam. The stripline flexible circuits installed in both receivers are capable of reading out 6 chains per circuit. Previous versions of the Mod-Cam readout harness utilized SMP-SMA transition PCBs to interface coax with the stripline; to reduce crosstalk we now directly connect SMP coax to the stripline. The Prime-Cam readout harness features identical designs, scaled for the additional circuits needed to read out multiple instrument modules.}
   \label{fig:readout_schematic}
\end{figure} 

KID arrays and their associated readout are mounted from the focal plane on the 100~mK stage, which is mechanically supported from the 1~K stage and thermally isolated via a roughly 12~cm tall low-thermal-conductivity carbon fiber truss \cite{vavagiakis_modcam_2022}. The 350~GHz module focal plane holds three detector arrays, each featuring six networks of detectors. Each detector network interfaces with a set of input and output coaxial cables connected via SMA connectors on the backside of the array packaging. Inside the module there are two temperature stage transitions for this coax between 100~mK--1~K and 1~K--4~K. 

In total the module contains 36 coax chains (18 pairs of inputs and outputs) that allow for the biasing of the KID detectors and readout of signals from the detector arrays. Input coax between 1~K--4~K uses low cost, robust 2.19~mm OD 304 Stainless Steel (SS), while the 100~mK-1~K transition uses thinner 1.19~mm OD 304 Stainless Steel for reduced thermal conductivity. To minimize thermal loading and compensate for the larger cross-sectional area of the cable, the 1~K--4~K transition utilizes nearly 0.5~m long input cables bent into a series of loops to maintain a compact form factor. The input chains also feature 10~dB attenuators mounted at 4~K, 6~dB attenuators at 1~K, and 6~dB attenuators at 100~mK. Both temperature transitions on the output coax use 2.19~mm OD superconducting Niobium-Titanium (NbTi) cables for favorable noise performance and low thermal conductivity up to the cryogenic amplifiers mounted at 4~K. A schematic of the module readout chains connecting to the readout harness installed in the receiver is shown in Fig. \ref{fig:readout_schematic}.

At the 4~K stage on the back of the instrument module, 18 cryogenic low noise amplifiers (LNAs) manufactured at Arizona State University amplify the output signals from the detectors and connect to the high density ``stripline" flexible readout circuits installed in both Prime-Cam and Mod-Cam that carry signals out of the receiver and to the warm Radio Frequency System on a Chip (RFSoC) readout electronics \cite{hamdi_amps, keller_stripline, sinclair_rfsoc}. The LNAs are mechanically mounted to a heatsink installed on the Cryoperm 4~K $\mathrm{\mu}$-metal magnetic shield that encloses the back of the module. The LNA heatsink is primarily cooled through a dedicated copper braid heat strap that mounts directly to the Prime-Cam 4~K plate. 

\subsection{Thermometry}

\begin{table*}[h]
    \centering
    \caption{Planned locations of thermometers to be installed in the 350~GHz module when it is deployed to the telescope for observations. The thermometers in the 280~GHz module are located in identical positions.}
    \vspace{0.125cm}
    \begin{tabular}{lll}
        \toprule
        \textbf{Location} & \textbf{Temperature} & \textbf{Type} \\
        \midrule
        Lens 1 & 4~K & Cernox 1050 \\
        Lens 2 & 1~K & ROX \\
        Focal Plane 1 & 100~mK & ROX \\
        Focal Plane 2 & 100~mK & ROX \\
        1~K Readout Plate & 1~K & ROX \\
        Low Noise Amplifier Heatsink & 4~K & Cernox 1050 \\
        \bottomrule
    \end{tabular}
    \label{tab:thermometers}
\end{table*}

The baseline deployment plan allocates six total thermometers monitoring critical thermal interfaces at the $100~\mathrm{mK}$, $1~\mathrm{K}$, and $4~\mathrm{K}$ stages for the $350~\mathrm{GHz}$ (and $280~\mathrm{GHz}$) modules in Prime-Cam. Cernox 1050 thin-film resistance cryogenic temperature sensors are used for monitoring $4~\mathrm{K}$ temperature stages while Ruthenium Oxide (ROX) thermometers are used for monitoring the $1~\mathrm{K}$ and $100~\mathrm{mK}$ stages. A summary of the locations of these thermometers is given in Table \ref{tab:thermometers}. To ensure redundancy for critical monitoring, two thermometers are allocated to the focal plane. The full $1~\mathrm{K}$ thermal path inside the module is bounded by the two $1~\mathrm{K}$ thermometers, while the 4~K path is bounded by the Lens 1 thermometer, along with other thermometers monitoring the temperature of the 4~K plate in Prime-Cam that are not listed here. The 4~K LNA heatsink thermometer is critical for estimating the amplifier noise contribution to readout noise in measured data. 

\subsection{Detector Arrays}
\label{sec:detectors}
\begin{figure}[t]
    \centering
    \includegraphics[width=0.8\linewidth]{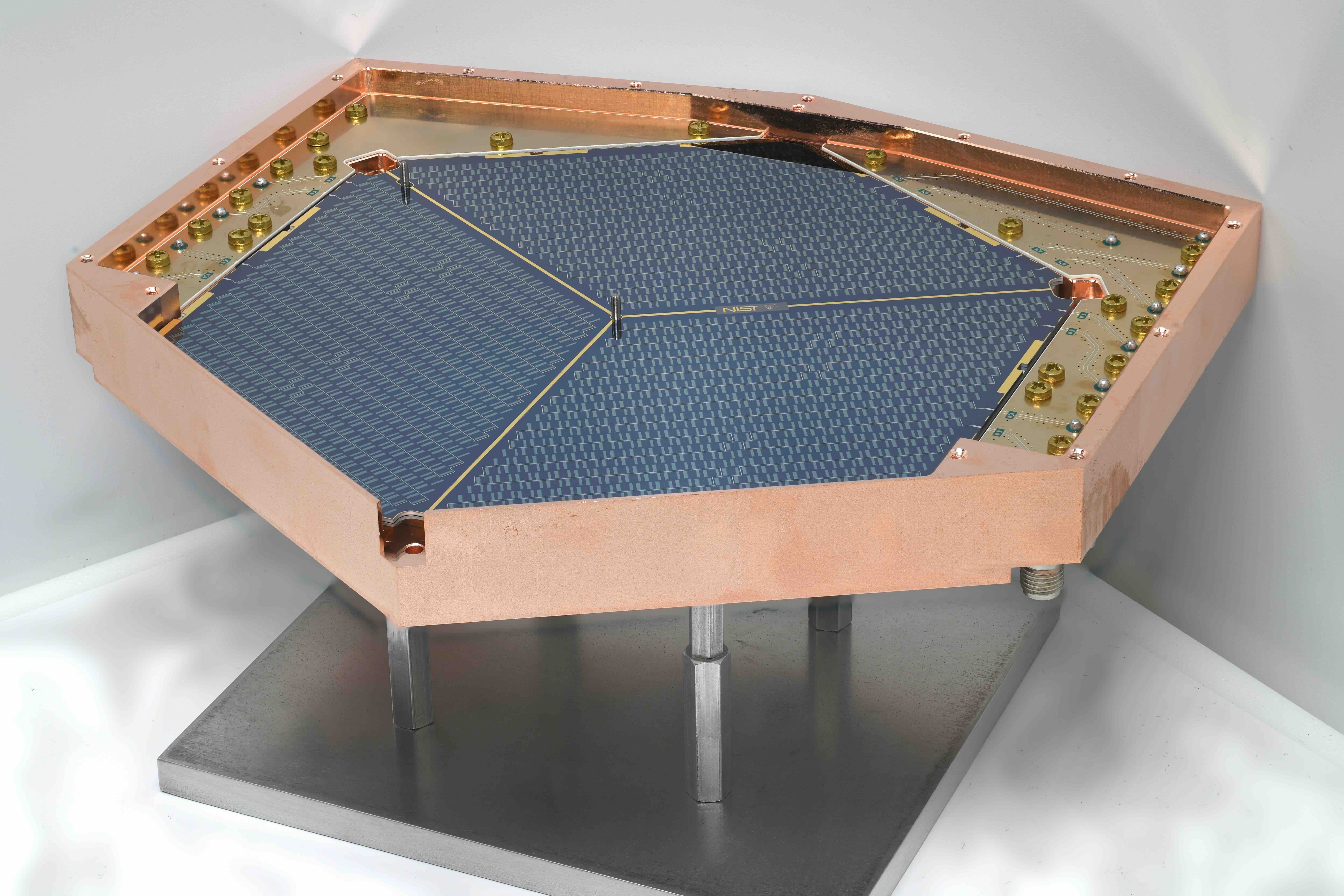}
    \caption{The first 350 GHz detector array shown in its packaging without the feedhorns installed.}
    \label{fig:array}
\end{figure}

\begin{figure}[b]
    \centering
    \includegraphics[width=0.6\linewidth]{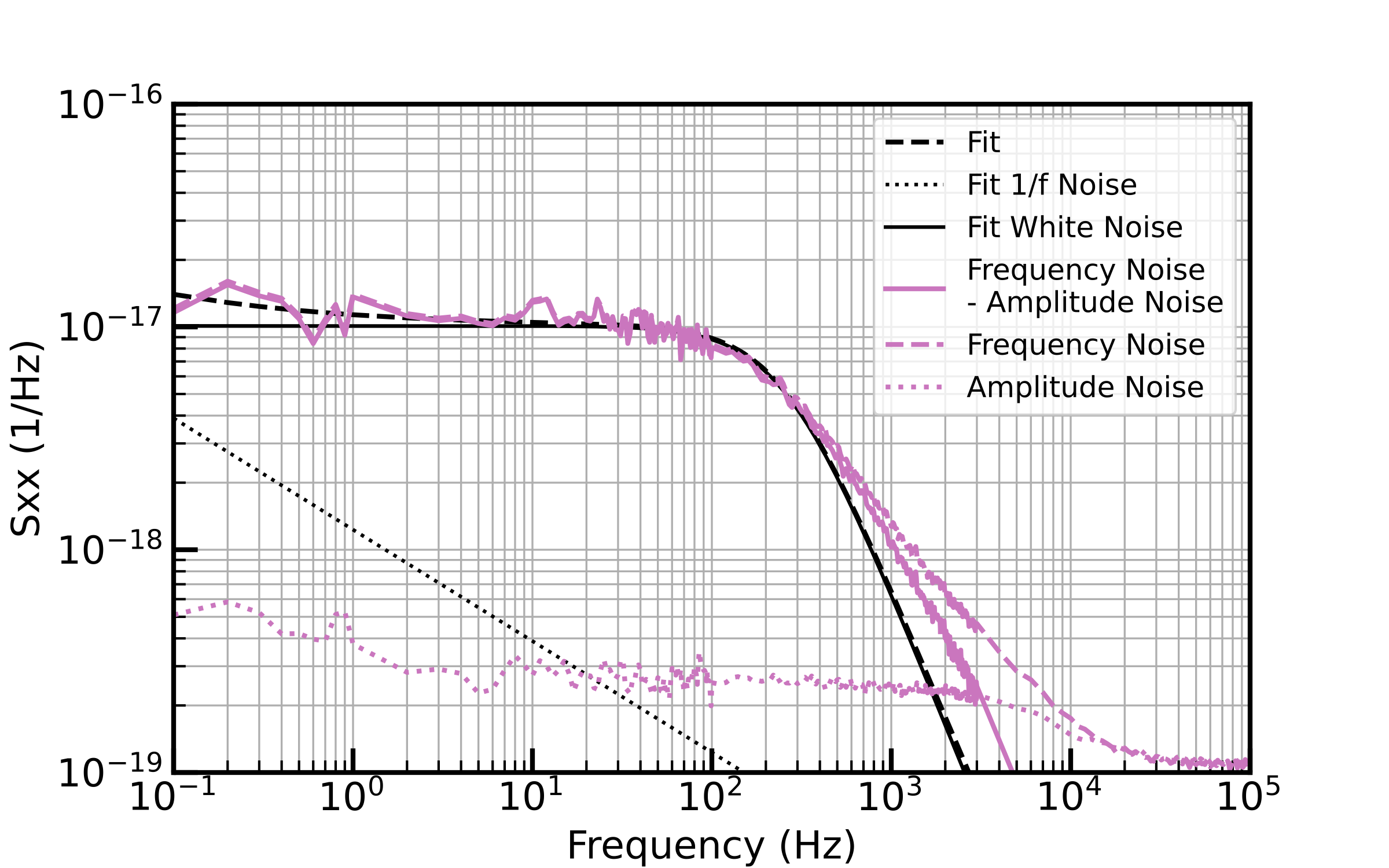}
    \caption{Noise power spectral density of a resonator on an Array 1 witness chip taken at 100 mK bath temperature with no optical loading. Noise is in Sxx units where x = df/f.  Even under no loading, the noise exhibits very little 1/f noise down to 0.1 Hz. }
    \label{Fig:noise}
\end{figure}

\begin{figure}[t]
    \begin{center}
    \includegraphics[width=1\linewidth]{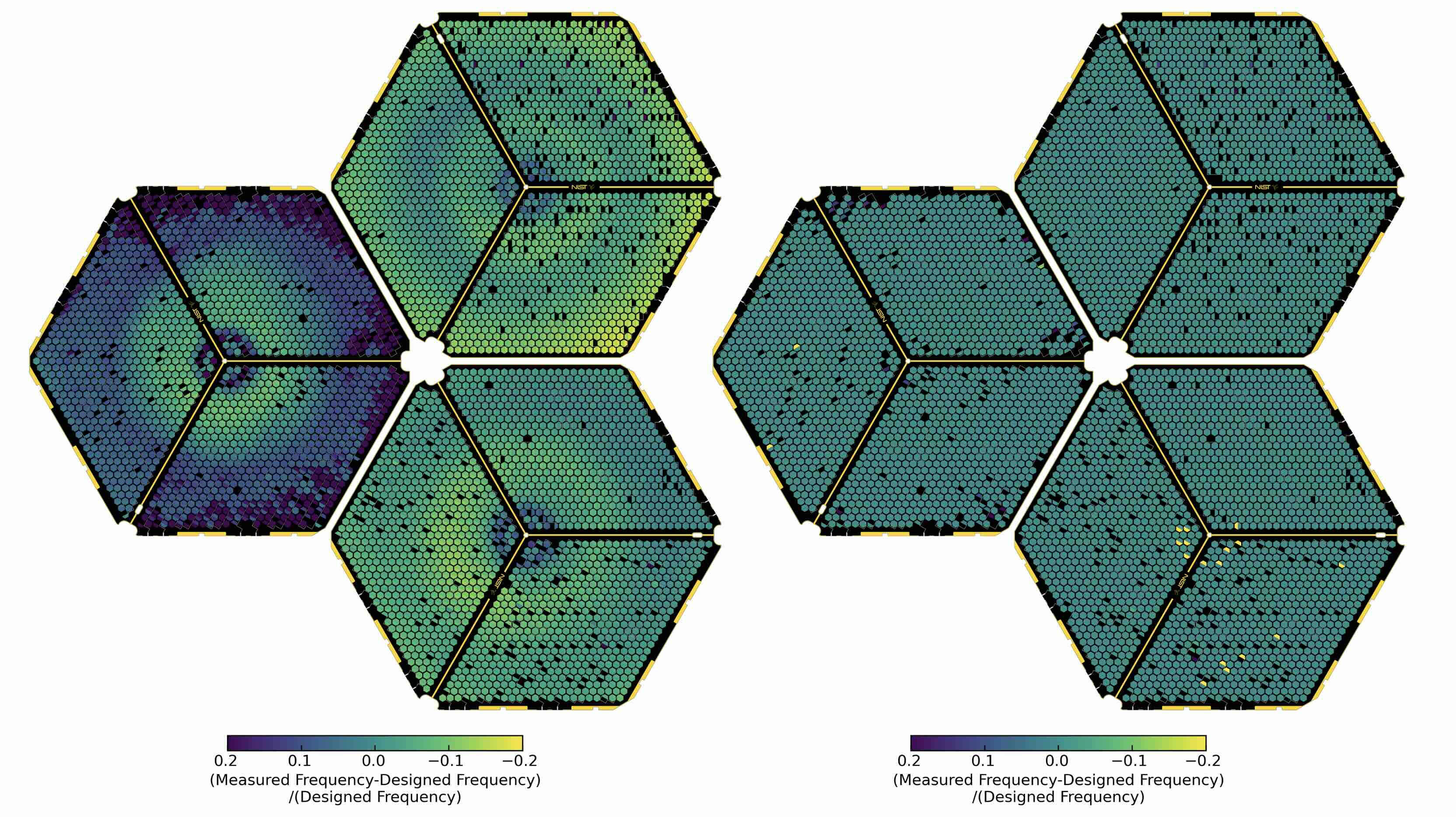}
    \end{center}
    \caption{Before-and-after capacitor editing: deviation of resonator frequency from the designed values. The left shows the fractional readout frequency deviation for all three arrays. Large radial deviation likely results from inductor linewidth variation across each wafer. After the resonator frequencies are measured via LED mapping, a new resonator-frequency scheduling design is developed by removing capacitor fingers from each resonator on each array. The right shows that the frequency targeting is greatly increased by this secondary editing, with less than a 1 percent deviation from the newly designed values.}
    \label{fig:350_arrays_trimmed} 
\end{figure}

\begin{figure}[htbp]
    \begin{center}

   \includegraphics[width=0.8\linewidth]{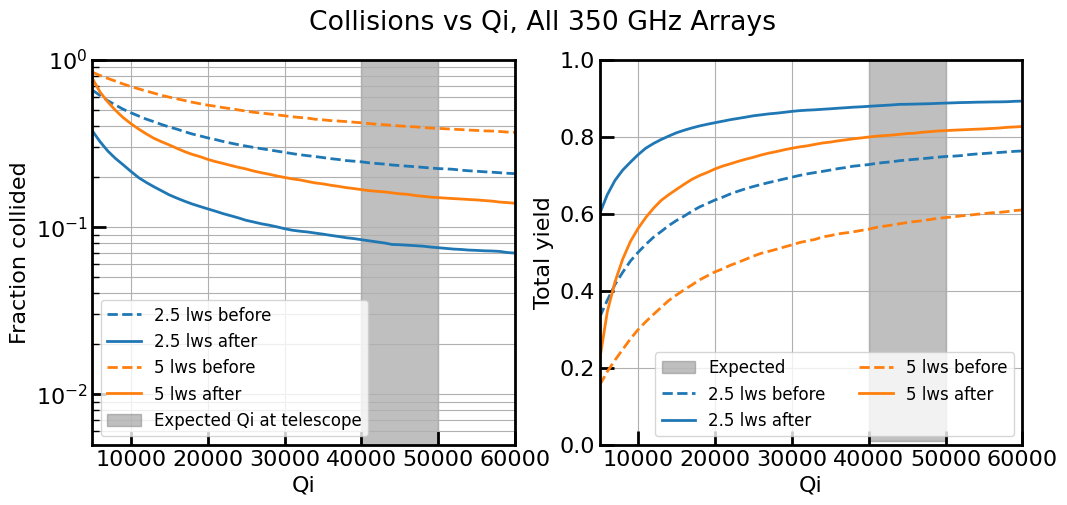}

   \end{center}
   \caption{
   \label{fig:NIST_lws}
    Results of capacitor trimming on all three 350 GHz arrays. Collisions depend on acceptable levels of crosstalk, which is approximately related to the resonator spacing in linewidths. The linewidth of each resonator depends on the telescope loading, which sets the internal quality factor Qi, and the as-made coupling quality factor Qc. The left panel shows the fraction of resonators within 2.5 and 5 linewidths of another resonator, before and after capacitor trimming. For the expected loaded Qi of 40,000 to 50,000, the collided fraction is reduced from 39\% to 15\% for a 5 linewidth separation and from 23\% to 8\% for a 2.5 linewidth separation. The right shows the total yield of uncollided resonators, including missing resonators. The five-linewidth yield increases from 58\% to 81\%  and the 2.5-linewidth yield increases from 74\% to 88\%. In terms of fractional increase, the 5-linewidth yield is boosted by a factor of 1.4, and the 2.5-linewidth yield is boosted by a factor of 1.19.}
\end{figure} 

\begin{table}[b]
\centering
\label{tab:resonator_yields}
\caption{Physical yield and loss analysis on the resonator frequency editing process.}
\vspace{0.125cm}
\begin{tabular}{ccccccc}
\toprule
 & \multicolumn{2}{c}{\textbf{Before Trimming}} & \multicolumn{2}{c}{\textbf{After Trimming}} & \multicolumn{2}{c}{\textbf{Trimming Loss}} \\
\midrule

\textbf{Array Number} & \textbf{Count} & \textbf{Yield (\%)} & \textbf{Count} & \textbf{Yield (\%)} & \textbf{Count} & \textbf{Loss (\%)} \\
\midrule

1     & 3317 / 3448 & 96.2 & 3305 / 3448 & 95.9 & 12 & 0.36 \\
2     & 3315 / 3448 & 96.1 & 3296 / 3448 & 95.6 & 19 & 0.57 \\
3     & 3310 / 3448 & 96.0 & 3306 / 3448 & 95.9 & 4  & 0.12 \\

\midrule
\textbf{Total} & \textbf{9942 / 10344} & \textbf{96.1} & \textbf{9907 / 10344} & \textbf{95.8} & \textbf{35} & \textbf{0.34} \\
\bottomrule
\end{tabular}
\end{table}

The 350~GHz instrument module has a focal plane populated with three arrays of polarization-sensitive, feedhorn-coupled Al KID detectors. Each array contains 3448 detectors made of 30~nm thick aluminum fabricated on a hexagonal 550~µm-thick, 15~cm diameter silicon-on-insulator (SOI) wafer. The device layer of the SOI is 63~µm thick and, with a backside deep etch and metallization, it forms a quarter-wave backshort for the KID absorber. One 350~GHz detector wafer is shown prior to feedhorn installation in Fig. \ref{fig:array}. The 350~GHz pixel design is based on previously fabricated 280~GHz arrays \cite{vaskuri_280ghz, duell_280ghz} drawing on BLAST-TNG and TOLTEC heritage\cite{blast_dets,toltec_dets}. The pixel pitch and detector counts remain the same as the 280 GHz aluminum arrays, along with the fabrication process. However, one important design distinction between the 280 GHz aluminum array and the 350 GHz arrays was a 50\% increase in capacitor area. This increase in area also allows for an increase in capacitor finger dimensions, allowing for the smallest finger dimension on the 280 GHz array of $6~\mathrm{\mu m}$ fingers with $6~\mathrm{\mu m}$ gaps to be increased to $10~\mathrm{\mu m}$ fingers with $10~\mathrm{\mu m}$ gaps. Increasing the capacitor size and finger dimensions can reduce TLS noise \cite{Noroozian_TLS}.
This design choice was made due to observed TLS noise in the capacitor of the 280~GHz detectors, resulting in a 0.5 Hz 1/f noise knee \cite{vaskuri_280ghz}. At first look, increasing the capacitor geometries appears to have reduced the 1/f noise; as shown in Fig. \ref{Fig:noise}, the noise measured on a witness chip fabricated alongside the array shows no significant 1/f noise down to 0.1 Hz even under dark conditions.

Due to limited frequency placement precision and typical resonator Qs of 30,000--100,000 under load, up to 40\% of KIDs in an array can end up within several linewidths of each other at typical multiplexing numbers.
This can be mitigated through tuning the resonant frequencies by physically editing the capacitors after the initial fabrication, allowing for a significant increase in resonator frequency placement precision.
Tuning requires mapping microwave resonant frequency to physical array location, accomplished here via LED mapping \cite{liu_ledmap,middleton_LEDmap}.
The increased frequency placement accuracy results in a substantial increase in the number of usable low cross talk detectors that is comparable to the gain that would be obtained by adding a fourth array, as is demonstrated in Figs. \ref{fig:350_arrays_trimmed} and \ref{fig:NIST_lws}.
All three 350 GHz arrays were successfully edited without issues with only a handful of resonators lost in the process (Tab. \ref{tab:resonator_yields}). 
This result supports that resonator frequency editing is a robust process that can scale to large focal planes of KIDs.

\begin{figure}[t]
    \centering
    \includegraphics[width=0.48\textwidth]{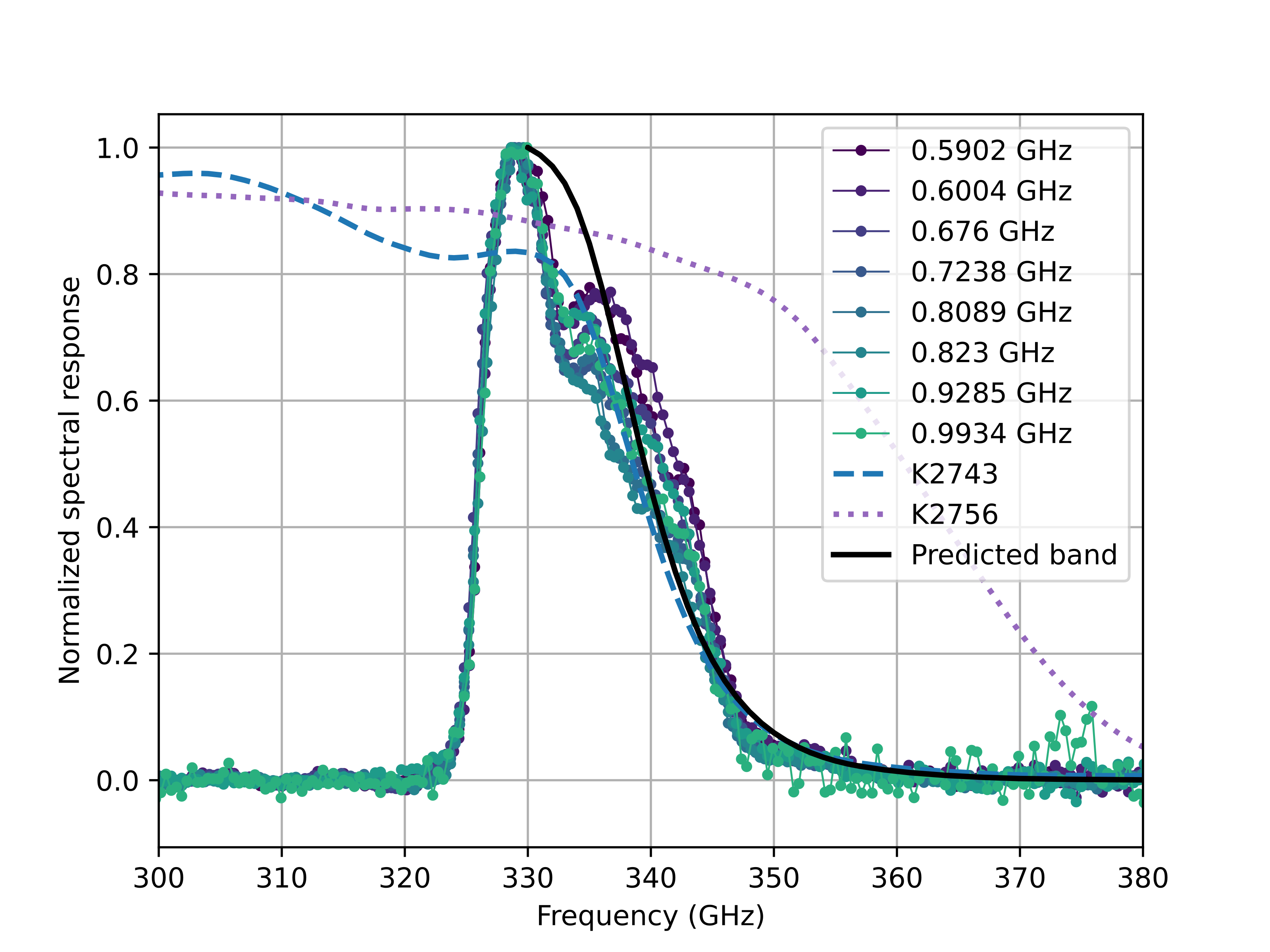}
    \hfill
    \includegraphics[width=0.48\textwidth]{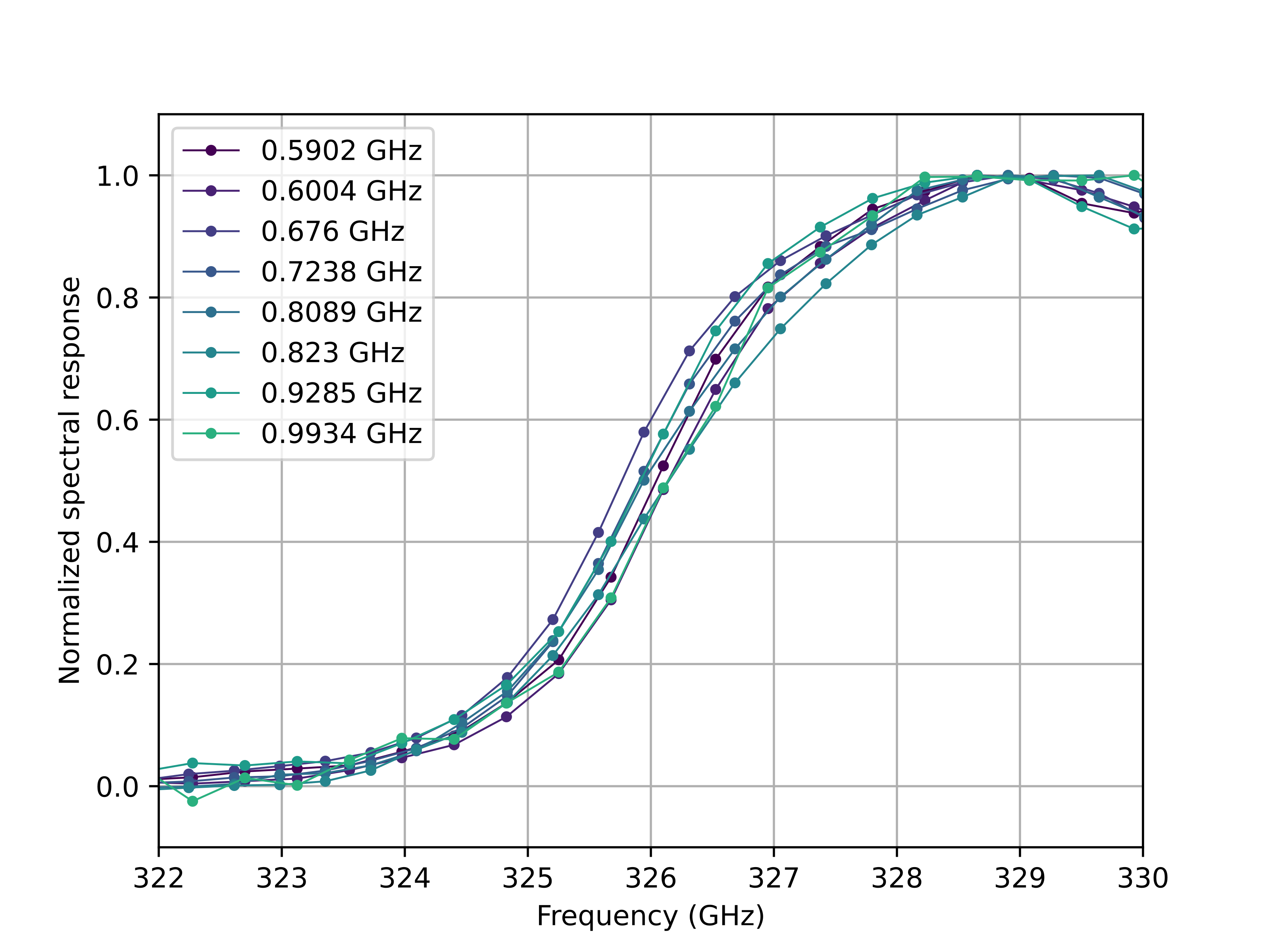}
    \caption{\textit{Left}: NIST FTS-measured passbands for a set of resonators taken with an arbitrary filter stack installed over Array 1. The transmission profiles of the two filters, $K2743$ and $K2756$ is overlaid. The full profile shown is not representative of the module passband due to the filters used, but the low-edge waveguide cutoff is expected to be similar. 
    \textit{Right}: A zoomed-in view of the lower edge of the passband from the NIST FTS measurement of Array 1. All measured detectors demonstrate good agreement with a 3~dB point around 326~GHz.}
    \label{fig:NIST_fts}
\end{figure}

Additionally, Fourier transform spectrometer (FTS) measurements performed at NIST on the first 350 GHz array (Array 1) with silicon platelet feedhorns installed have been used to characterize the low edge of the detector array passband set by the feedhorn waveguide cutoff. For this test, filters differing from the nominal in-module filters described in Section \ref{sec:optics} were used, hence, the high edge of the passband should not be considered representative of the actual module passband edge. The passbands for a select group of detectors were measured and are shown in Fig. \ref{fig:NIST_fts}. Results demonstrate a low edge 3~dB point of approximately 326~GHz for this array, which confirms direct room-temperature measurements of the feedhorn waveguide cutoff taken with a millimeter-wave vector network analyzer \cite{austermann_feeds}. Subsequent arrays had feedhorns receive an additional round of gold plating in order to reduce the diameter of the waveguide aperture, thereby shifting the waveguide cutoff to a higher frequency.

\section{Instrument Module Testing}
\label{sec:testing}

Testing of the full 350~GHz module was performed at Cornell University with the module ``optically open" to the room in a configuration that would enable cryogenic validation in an environment similar to that of Prime-Cam. This configuration follows the deployment configuration with two exceptions: 1) the baseline Lyot stop at the module pupil was replaced with a Lyot stop with 1/3 of the aperture size to reduce optical loading from the room on the detectors, and 2) the UHMWPE window on the vacuum shell was replaced with a double-thickness UHMWPE window to compensate for the nearly two times greater ambient pressure in the lab. The smaller aperture Lyot stop is expected to reduce loading to levels more similar to that expected on-sky, but may cause vignetting for pixels near the focal plane edge. However, we only characterize detectors located near the center of the focal plane in this work. The double-thickness UHMWPE window meanwhile lacks an antireflective coating and may therefore contribute a small sinusoidal component to the measured optical transmission through the module due to its low index of refraction.

\subsection{Cryogenic Validation}

\begin{figure}[htbp]
    \centering
    \includegraphics[width=0.8\textwidth]{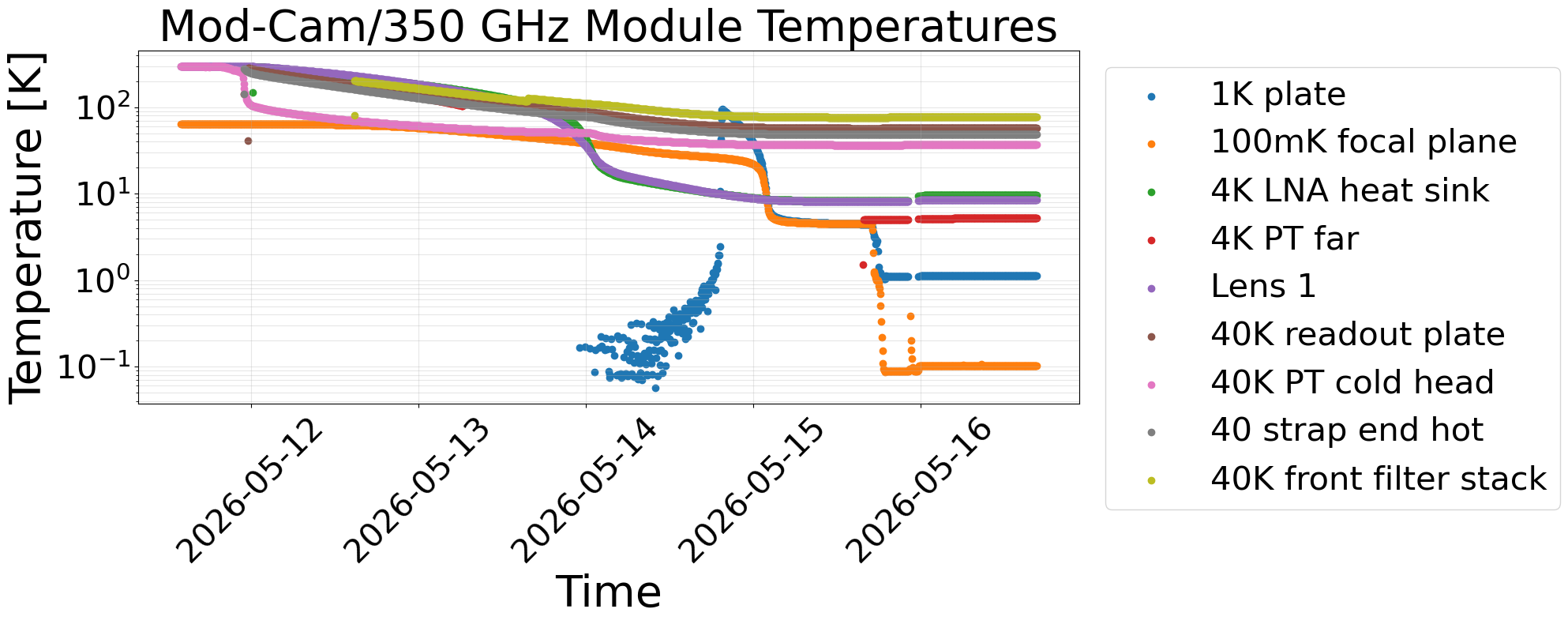}
    
    \vspace{0.1cm}
    
    \includegraphics[width=0.8\textwidth]{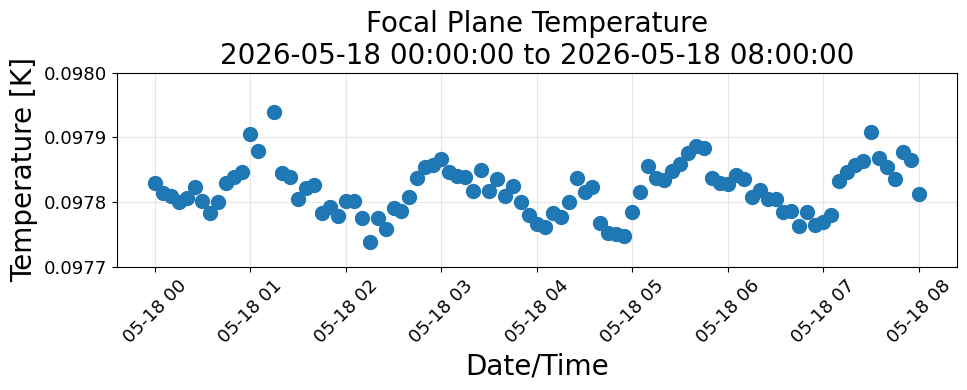}
    
    \caption{
    \textit{Top:} Mod-Cam and 350~GHz module 40~K-and-under temperatures are logged as a function of time during the cooldown period. The condensing cycle rapidly drops the 100~mK and 1~K stage temperatures and is begun midway through the day on 2026-05-15. 
    \textit{Bottom:} Temperature variations on the focal plane during an ``optically quiet" period when no specific optical or detector tests were performed. During this period amplifiers and detectors remained biased but were not streaming data. The low sampling rate reflects the scanning of the Lakeshore LS372 over multiple channels monitoring all Mod-Cam and 350~GHz module thermometers during this period. The difference between this thermometer temperature and the intended 100~mK setpoint represents a minor temperature gradient across the focal plane not expected to significantly impact operations given the demonstrated thermal stability.      }
    \label{fig:cryo_performance}
\end{figure}


The 350~GHz module installed in Mod-Cam took approximately 3.5 days after turning on the Mod-Cam pulse tubes to cool to 4~K base temperatures, at which point the dilution refrigerator (DR) condensing cycle could begin. Condensing the DR to reach stable 100~mK base temperatures then took an additional 1 hour. A plot of the module and sub-40~K Mod-Cam temperatures during the cooldown is shown in Fig. \ref{fig:cryo_performance}. Though some 4~K and 40~K temperature stages in \ref{fig:cryo_performance} show elevated base temperatures, this is a known feature of the Mod-Cam cryogenic environment that did not meaningfully impact testing. Furthermore, additional cooling power in Prime-Cam has been shown sufficient to fully cool these stages to their expected base temperatures \cite{vavagiakis_pcam_inprep}.

The time required to cool the 350~GHz module in Mod-Cam is consistent with prior published results for the 280~GHz module in Mod-Cam and models for the cryogenic behavior of the receiver \cite{lin2025ccatmodcamcryogenicperformance}\cite{gascard_thermal_model}. These metrics for both instrument modules in Mod-Cam are also roughly equivalent to the time to reach base temperatures in Prime-Cam \cite{vavagiakis_pcam_inprep}.  

In the installed configuration, the module achieved stable 80~mK focal plane base temperatures in Mod-Cam with only moderate levels of DR still power optimization, leaving room for further improvement. The focal plane was then servoed up to 100~mK to provide cryogenic overhead to keep focal plane temperatures constant for long time periods even when biasing detectors. 100~mK is the baseline operating temperature for the focal planes of modules installed in Prime-Cam. The 350~GHz module demonstrated stable cryogenic temperatures during testing for timescales \textgreater 1 week, during which time tests were performed to characterize the module optics and detectors. During this time there were no indications of transient thermal effects or degradations in cooling power.


We can quantitatively assess focal plane stability over long periods by selecting an 8 hour timestream of temperature data during a period when no optical testing was performed. We then calculate the root mean square focal plane temperature variation, $\Delta T_{RMS}$, finding $\Delta T_{RMS} = 3.9\times 10^{-5} ~\mathrm{K}$. The focal plane temperatures during this period are shown in the right panel of Fig. \ref{fig:cryo_performance}. This result is within 20\% of the variations seen in the 280~GHz module focal plane over similar timescales, albeit in a more stable dark optical environment \cite{lin2025ccatmodcamcryogenicperformance}. Regardless, at this magnitude of fluctuation, focal plane temperature variations are expected to generate only trivial fractional frequency shifts in our detectors compared to shifts induced by on-sky loading at the CCAT site \cite{lin2025ccatmodcamcryogenicperformance}. 

\subsubsection{Yield}

\begin{figure}[t]
    \centering
    \includegraphics[width=0.9\linewidth]{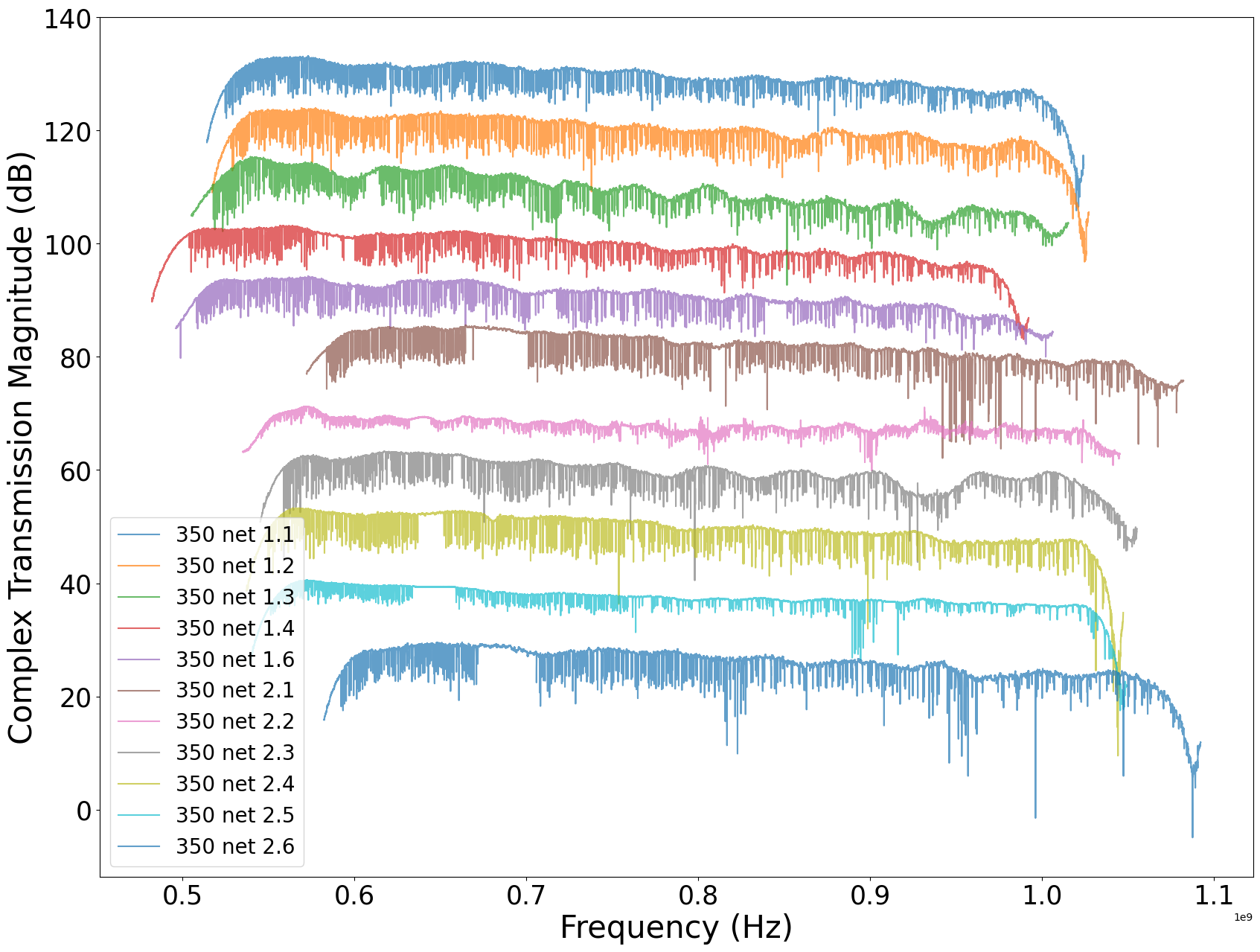}
    \caption{VNA sweeps for each detector network showing all resonators read out during the testing cooldown. The complex transmission magnitude shown is arbitrary, and is offset for each network to increase visibility between the traces. Array 1 Network 5 (``Net 1.5", not shown) suffered a broken LNA and was unable to be read out during this cooldown.}
    \label{Fig:vna_sweep}
\end{figure}

Preliminary detector yield estimates were taken on 11/12 functional networks across the two detector arrays installed for this cooldown. During the cooldown, one network suffered a broken amplifier solder joint that rendered the amplifier unusable. Across the 11 networks we successfully found \textgreater 82\% of designed resonators with the default parameter settings assigned to our peakfinding algorithm, yielding a total of more than 5200 found resonators. VNA sweeps for all functional networks are shown in Fig. \ref{Fig:vna_sweep}. This yield estimate is roughly consistent with the percentage of resonators separated by \textgreater 5 linewidths after trimming under expected on-sky loading conditions shown in Fig. \ref{fig:NIST_fts}, however further work is needed to identify the true fraction of collided resonators in these loading conditions.

\subsubsection{Detector Mapping}
\label{sec:det_mapping}

\begin{figure}[t]
   \begin{center}
   \includegraphics[width=1\linewidth]{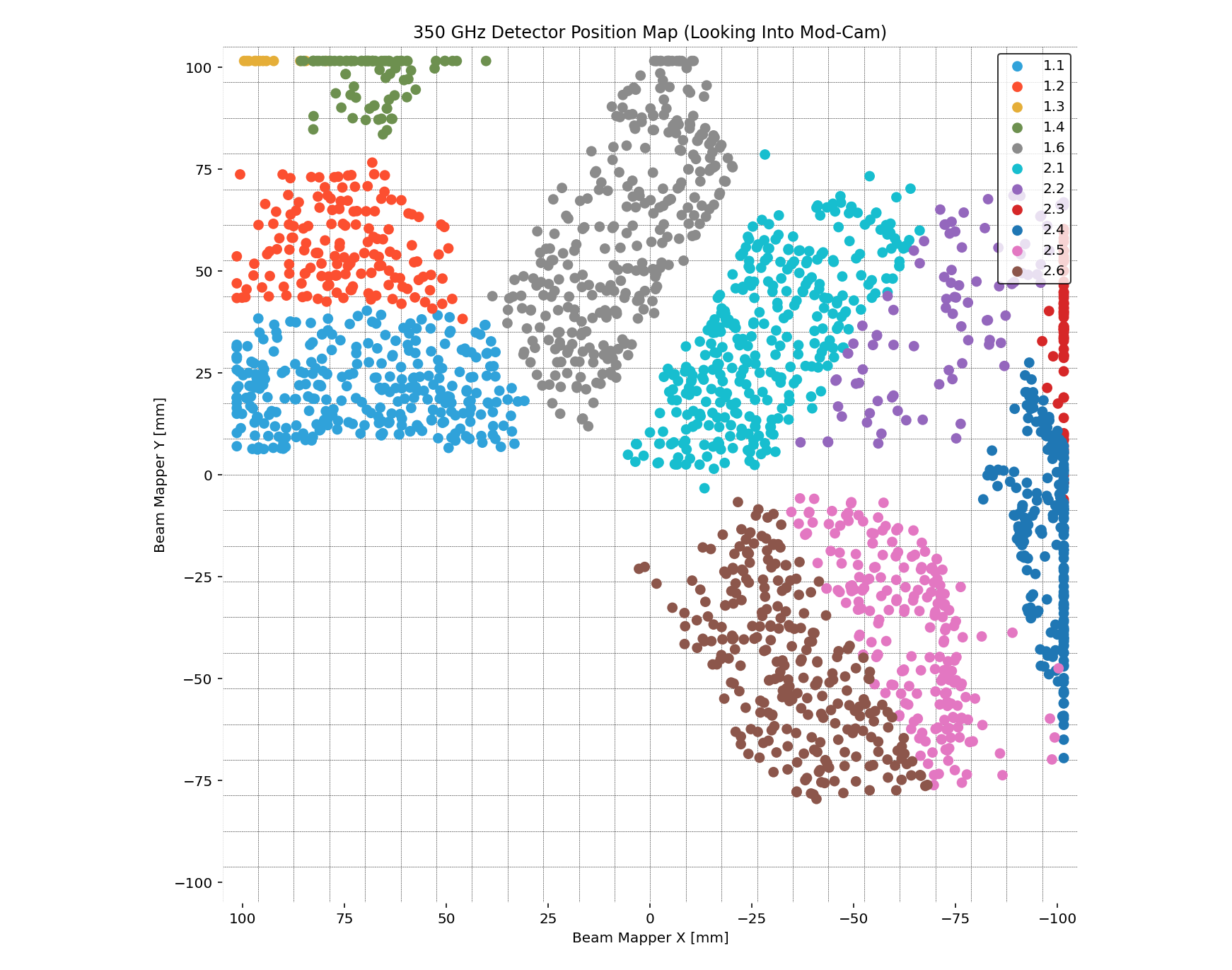}
   \end{center}
   \caption{Preliminary inferred resonator positions measured across both detector arrays. Poorly localized or unphysical resonators have been cut from this plot for clarity. Resonators falling near the edge of the mapped region do not have a well-constrained position due to the inability of the 2D Gaussian fit to identify a reasonable near-field beam pattern. Only a short amount of time was allocated for these measurements so that the module could be packed in time to ship it to Chile with Prime-Cam.}
   \label{fig:det_map} 
\end{figure} 
   
In dark environments with low loading, LED mapping\cite{middleton_LEDmap} as described in Section \ref{sec:detectors} may be used to create a map of resonator frequencies to physical positions by individually illuminating pixels and identifying corresponding resonators with high accuracy, facilitating capacitor trimming to increase yields by standardizing resonator spacing. However, in more realistic operation of the instrument we cannot have LEDs obstructing the optical path, and increased loading from optical coupling may also cause some fraction of resonators to collide or cross each other in frequency space relative to the LED-mapper result. As a result, we need methods to construct a mapping between resonators and physical pixel locations during realistic operation of the instrument, whether in-lab or on-sky.  

To create a mapping of resonators to physical positions we installed a ``beam mapper” in front of the Mod-Cam window consisting of a Hawkeye IR-55 source mounted to a translating x-y stage capable of moving across a 20~cm x 20~cm square region \cite{patel2025ccatreadout}. The translating stage used black carbon-loaded cloth covering reflective regions to eliminate spurious optical signals not from the source itself. The source has a collimating optic installed with a beam full width at half maximum (FWHM) of 15 degrees. 

The beam mapper source was flashed at 10~Hz while the stage is stepped across the full 21~cm x 21~cm square in 7~mm increments. When movement is paused at each position of the beam mapper, a full set of timestreams is taken across all detector networks. We then fourier transform the measured timestream for each detector and subtract the detector white noise measured in a high frequency region far above the 1/f knee and signal frequency. We then take the peak height at 10 Hz as the signal-to-noise ratio of the flashing source, reporting this as the signal amplitude at the given beam mapper position. The process is repeated for the entire 21~cm x 21~cm grid. 

The resulting grid of measured signal amplitudes generates a near-field beam profile for each detector due to our lack of focusing optics. The beams are fit with a symmetric 2D-gaussian to identify the beam center, which is assumed to correspond with the position of the detector feedhorn. The reconstruction of all beam centers for all detectors allows us to build the mapping of resonators to physical positions on the focal plane, as is shown in Fig. \ref{fig:det_map}. Because the source is unfocused and appears extended in this configuration, jitter introduced by wide beam profiles makes it difficult to identify exact pixels for each detector, though some grid-like patterns we might expect for our designed array geometry are apparent in Array 2 Network 1, for example. We expect when the instrument is coupled to the telescope, the ability to view true point sources, including planets such as Uranus, through the focusing optics should improve our ability to locate resonators when a similar procedure is applied to reconstruct the detector mapping on-sky.  

\subsubsection{Passband Measurement}

   \begin{figure}
   \begin{center}
   \includegraphics[width=1\linewidth]{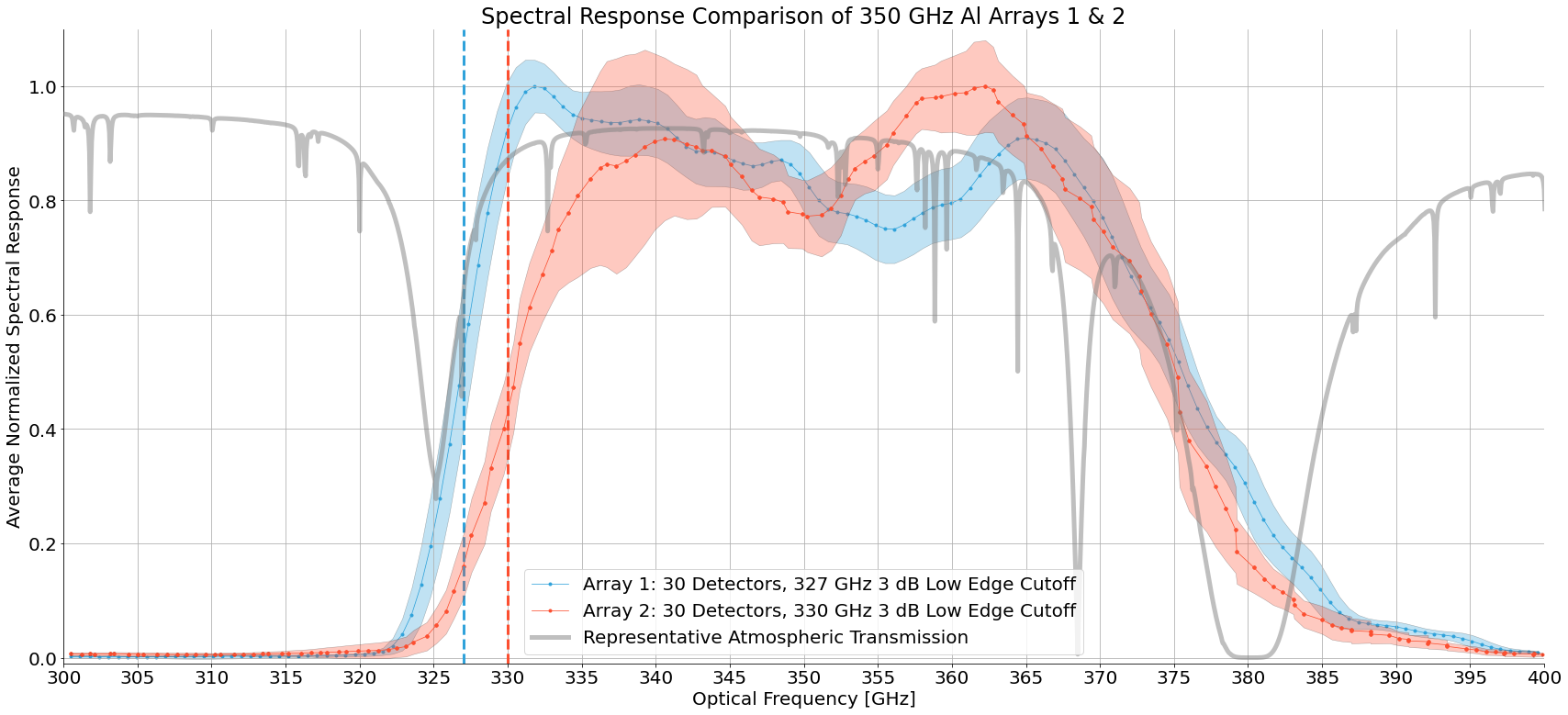}
   \end{center}
   \caption[test] 
   { \label{fig:cornell_fts_bandpass} 
The averaged passband for the 30 highest signal-to-noise ratio detectors in each of the first two 350~GHz arrays is shown. The gray line overlaid shows a normalized representative atmospheric transmission in this band calculated with the \textit{am} atmospheric model \cite{amcode_model}. }
   \end{figure} 

We perform passband measurements for the 350 GHz module in Mod-Cam using a Martin-Puplett Fourier Transform Spectrometer in a Mach-Zehnder configuration based on the PIXIE FTS design\cite{pan_pixie, pixie_design}, as has been previously used for characterizing the 280~GHz module passband \cite{patel2025ccatreadout, vavagiakis_pcam_inprep}. The measurement was taken with Mod-Cam optically open to the room and all lenses, lowpass filters, and IR-blocking filters described in Section \ref{sec:optics} installed. The aim of this test was to characterize the full module passband in a configuration close to that which we plan to deploy at the telescope. 

Our FTS utilizes a $\sim700$ K blackbody as a broadband source at one input and a room-temperature source at the other \cite{patel2025ccatreadout}. After tuning the detectors, we measure interferograms by recording timestreams from all detector networks while the optical path delay is varied by translating the central mirror of the FTS at a speed of 1 mm/s, up to a maximum displacement of 7~cm from the central position. The spectral response is then calculated by using the timestamped positions of the mirror $y$. The optical delay $d$ is given by $d \approx 4y$ for small mirror displacements \cite{pan_pixie}. The resonator fractional frequency shift vs. optical delay interferograms were then Fourier transformed to obtain detector spectral responses. 

The FTS was positioned for our measurement so that the output beam entered perpendicular to the Mod-Cam window. FTS alignment errors are expected to be the dominant source of passband measurement systematics and could shift the entire measured passband, so a routine was developed to ensure good alignment. By using the signal-to-noise ratio of each detector's spectral response assigned to the location derived from the detector maps, we were able to check the FTS illumination pattern across the focal plane for uniformity and position. This allowed us to iterate on and fine tune the FTS positioning until good alignment was demonstrated. Furthermore, this routine also confirmed we were measuring each array's passband individually, as intended.

Fig. \ref{fig:cornell_fts_bandpass} shows the averaged spectral responses for the 30 highest signal-to-noise ratio detectors in each of the measured 350 GHz KID arrays. For both arrays, the selected detectors lie physically central to the focal plane and are located within the center of the FTS illumination pattern, in order to reduce systematic effects on the passband introduced by partial FTS illumination and alignment. Both arrays demonstrate a band center of approximately 350~GHz as expected; for Array 1 the low cutoff 3~dB point occurs at 327 GHz while for Array 2 the low cutoff 3~dB point occurs at 330 GHz. This result confirms measurements taken at NIST using a mm-wave VNA that demonstrate the ability to shift the feedhorn waveguide cutoff higher post-fabrication with additional gold plating, as was applied to the feedhorns of Array 2 \cite{austermann_feeds}. The difference between the low-edge cutoff for Array 1 and the corresponding cutoff measured at NIST and shown in Fig. \ref{fig:NIST_fts} is also within the expected uncertainty for our FTS. For both arrays, the high edge of the band, defined by the shared 100~mK lowpass edge filter, has a 3~dB point of 375 GHz.

\section{Conclusion}
The 350~GHz broadband instrument module has been successfully tested in Mod-Cam and is scheduled for deployment in 2026. The module demonstrates stable cryogenic operation over long timescales with high yield in the detector arrays under realistic loading conditions. We have also shown the ability to physically locate detectors using a translating source, which will become critical during early on-sky commissioning of the instrument in order to reconstruct per-detector pointings. The full module passband measurements have confirmed that re-plating the feedhorns with additional gold is a viable option for increasing the waveguide cutoff and avoiding the atmospheric water line around 320~GHz that would otherwise result in mapping speed degradation \cite{austermann_feeds}. On the high edge of the measured passband, we find the cutoff at 375~GHz extends in to the water line where little atmospheric transmission occurs, as shown on the right side of Fig. \ref{fig:cornell_fts_bandpass}. Efforts are currently underway to fabricate a new deployment-grade 100~mK LPE with a target cutoff of 364~GHz.

\appendix    

\acknowledgments 
 
The construction of the 350 GHz instrument module for Prime-Cam was supported by NSF grant AST-2117631. 

The CCAT project, FYST and Prime-Cam instrument have been supported by generous contributions from the Fred M. Young, Jr. Charitable Trust, Cornell University, Duke University, and the Canada Foundation for Innovation and the Provinces of Ontario, Alberta, and British Columbia. The construction of the FYST telescope was supported by the Gro{\ss}ger{\"a}te-Programm of the German Science Foundation (Deutsche Forschungsgemeinschaft, DFG) under grant INST 216/733-1 FUGG, as well as funding from Universit{\"a}t zu K{\"o}ln, Universit{\"a}t Bonn, and the Max Planck Institut f{\"u}r Astrophysik, Garching. 

The completion and deployment of the Prime-Cam instrument with the initial instrument modules is supported by a generous contribution from Alex Gerko, Founder and CEO of XTX Markets.

\bibliography{report} 

@ARTICLE{ccatsciencepaper,
       author = {{CCAT Collaboration} and {Aravena}, Manuel and {Austermann}, Jason E. and {Basu}, Kaustuv and {Battaglia}, Nicholas and {Beringue}, Benjamin and {Bertoldi}, Frank and {Bigiel}, Frank and {Bond}, J. Richard and {Breysse}, Patrick C. and {Broughton}, Colton and {Bustos}, Ricardo and {Chapman}, Scott C. and {Charmetant}, Maude and {Choi}, Steve K. and {Chung}, Dongwoo T. and {Clark}, Susan E. and {Cothard}, Nicholas F. and {Crites}, Abigail T. and {Dev}, Ankur and {Douglas}, Kaela and {Duell}, Cody J. and {D{\"u}nner}, Rolando and {Ebina}, Haruki and {Erler}, Jens and {Fich}, Michel and {Fissel}, Laura M. and {Foreman}, Simon and {Freundt}, R.~G. and {Gallardo}, Patricio A. and {Gao}, Jiansong and {Garc{\'\i}a}, Pablo and {Giovanelli}, Riccardo and {Golec}, Joseph E. and {Groppi}, Christopher E. and {Haynes}, Martha P. and {Henke}, Douglas and {Hensley}, Brandon and {Herter}, Terry and {Higgins}, Ronan and {Hlo{\v{z}}ek}, Ren{\'e}e and {Huber}, Anthony and {Huber}, Zachary and {Hubmayr}, Johannes and {Jackson}, Rebecca and {Johnstone}, Douglas and {Karoumpis}, Christos and {Keating}, Laura C. and {Komatsu}, Eiichiro and {Li}, Yaqiong and {Magnelli}, Benjamin and {Matthews}, Brenda C. and {Mauskopf}, Philip D. and {McMahon}, Jeffrey J. and {Meerburg}, P. Daniel and {Meyers}, Joel and {Muralidhara}, Vyoma and {Murray}, Norman W. and {Niemack}, Michael D. and {Nikola}, Thomas and {Okada}, Yoko and {Puddu}, Roberto and {Riechers}, Dominik A. and {Rosolowsky}, Erik and {Rossi}, Kayla and {Rotermund}, Kaja and {Roy}, Anirban and {Sadavoy}, Sarah I. and {Schaaf}, Reinhold and {Schilke}, Peter and {Scott}, Douglas and {Simon}, Robert and {Sinclair}, Adrian K. and {Sivakoff}, Gregory R. and {Stacey}, Gordon J. and {Stutz}, Amelia M. and {Stutzki}, Juergen and {Tahani}, Mehrnoosh and {Thanjavur}, Karun and {Timmermann}, Ralf A. and {Ullom}, Joel N. and {van Engelen}, Alexander and {Vavagiakis}, Eve M. and {Vissers}, Michael R. and {Wheeler}, Jordan D. and {White}, Simon D.~M. and {Zhu}, Yijie and {Zou}, Bugao},
        title = "{CCAT-prime Collaboration: Science Goals and Forecasts with Prime-Cam on the Fred Young Submillimeter Telescope}",
      journal = {Apjs},
         year = 2023,
        month = jan,
       volume = {264},
       number = {1},
          eid = {7},
        pages = {7},
          doi = {10.3847/1538-4365/ac9838},
archivePrefix = {arXiv},
       eprint = {2107.10364},
 primaryClass = {astro-ph.CO},
       adsurl = {https://ui.adsabs.harvard.edu/abs/2023ApJS..264....7C}
}

@inproceedings{Parshley_2018,
   title={CCAT-prime: a novel telescope for sub-millimeter astronomy},
   url={http://dx.doi.org/10.1117/12.2314046},
   DOI={10.1117/12.2314046},
   booktitle={Ground-based and Airborne Telescopes VII},
   publisher={SPIE},
   author={Parshley, Stephen C. and Kronshage, Jörg and Bazarko, Andrew and Bertoldi, Frank and Bustos, Ricardo and Campbell, Donald B. and Chapman, Scott and Cothard, Nicolas and Devlin, Mark and Erler, Jens and Fich, Michel and Gallardo, Patricio A. and Giovanelli, Riccardo and Graf, Urs and Gramke, Scott and Haynes, Martha P. and Blair, James and Herter, Terry and Nolta, Mike and Stacey, Gordon J. and Hills, Richard and Limon, Michele and Mangum, Jeffrey G. and McMahon, Jeff and Niemack, Michael D. and Nikola, Thomas and Omlor, Markus and Riechers, Dominik A. and Steeger, Karl and Stutzki, Juergen and Vavagiakis, Eve M.},
   editor={Gilmozzi, Roberto and Marshall, Heather K. and Spyromilio, Jason},
   year={2018},
   month=July, pages={220} }

@inproceedings{vavagiakis_modcam_2022,
	author = {Eve M. Vavagiakis and Cody J. Duell and Jason Austermann and James Beall and Tanay Bhandarkar and Scott C. Chapman and Steve K. Choi and Gabriele Coppi and Simon Dicker and Mark Devlin and Rodrigo G. Freundt and Jiansong Gao and Christopher Groppi and Terry L. Herter and Zachary B. Huber and Johannes Hubmayr and Doug Johnstone and Ben Keller and Anna M. Kofman and Yaqiong Li and Philip Mauskopf and Jeff McMahon and Jenna Moore and Colin C. Murphy and Michael D. Niemack and Thomas Nikola and John Orlowski-Scherer and Kayla M. Rossi and Adrian K. Sinclair and Gordon J. Stacey and Joel Ullom and Michael Vissers and Jordan Wheeler and Zhilei Xu and Ningfeng Zhu and Bugao Zou},
	booktitle = {Millimeter, Submillimeter, and Far-Infrared Detectors and Instrumentation for Astronomy XI},
	editor = {Jonas Zmuidzinas and Jian-Rong Gao},
	organization = {International Society for Optics and Photonics},
	pages = {1219004},
	publisher = {SPIE},
	title = {{CCAT-prime: design of the Mod-Cam receiver and 280 GHz MKID instrument module}},
	volume = {12190},
	year = {2022}}

@misc{patel2025ccatreadout,
      title={CCAT: Readout of over 10,000 280 GHz KIDs in Mod-Cam using RFSoC Electronics}, 
      author={Darshan A. Patel and Yuhan Wang and Cody J. Duell and Jason E. Austermann and James Beall and James R. Burgoyne and Scott Chapman and Steve K. Choi and Rodrigo G. Freundt and Eliza Gazda and Christopher Groppi and Zachary B. Huber and Johannes Hubmayr and Ben Keller and Lawerence T. Lin and Philip Mauskopf and Alicia Middleton and Michael D. Niemack and Cody Roberson and Adrian K. Sinclair and Ema Smith and Jeff van Lanen and Anna Vaskuri and Benjamin J. Vaughan and Eve M. Vavagiakis and Michael Vissers and Samantha Walker and Jordan Wheeler and Ruixuan and Xie},
      year={2025},
      eprint={2510.06491},
      archivePrefix={arXiv},
      primaryClass={astro-ph.IM},
      url={https://arxiv.org/abs/2510.06491}, 
}

@misc{lin2025ccatmodcamcryogenicperformance,
      title={CCAT: Mod-Cam Cryogenic Performance and its Impact on 280 GHz KID Array Noise}, 
      author={Lawrence T. Lin and Eve M. Vavagiakis and Jason E. Austermann and James R. Burgoyne and Scott Chapman and Steve K. Choi and Abigail T. Crites and Cody J. Duell and Rodrigo G. Freundt and Eliza Gazda and Christopher Groppi and Anthony I. Huber and Zachary B. Huber and Johannes Hubmayr and Ben Keller and Philip Mauskopf and Alicia Middleton and Michael D. Niemack and Darshan A. Patel and Cody Roberson and Adrian K. Sinclair and Ema Smith and Anna Vaskuri and Benjamin J. Vaughan and Samantha Walker and Yi Wang and Yuhan Wang and Jordan Wheeler and Ruixuan and Xie},
      year={2025},
      eprint={2509.25018},
      archivePrefix={arXiv},
      primaryClass={astro-ph.IM},
      url={https://arxiv.org/abs/2509.25018}, 
}

@article{Sierra_2025_SO_optics_tube,
	author = {Sierra, Carlos E. and Harrington, Kathleen and Sutariya, Shreya and Alford, Thomas and Kofman, Anna M. and Chesmore, Grace E. and Austermann, Jason E. and Bazarko, Andrew and Beall, James A. and Bhandarkar, Tanay and Devlin, Mark J. and Dicker, Simon R. and Dow, Peter N. and Duff, Shannon M. and Dutcher, Daniel and Galitzki, Nicholas and Golec, Joseph E. and Groh, John C. and Gudmundsson, Jon E. and Haridas, Saianeesh K. and Healy, Erin and Hubmayr, Johannes and Iuliano, Jeffrey and Johnson, Bradley R. and Lessler, Claire S. and Lew, Richard A. and Link, Michael J. and Lucas, Tammy J. and McMahon, Jeffrey J. and Moore, Jenna E. and Nati, Federico and Niemack, Michael D. and Schmitt, Benjamin L. and Silva-Feaver, Max and Singh, Robinjeet and Sonka, Rita F. and Thomas, Alex and Thornton, Robert J. and Tsan, Tran and Ullom, Joel N. and Van Lanen, Jeffrey L. and Vavagiakis, Eve M. and Vissers, Michael R. and Wang, Yuhan and Zheng, Kaiwen},
	journal = {The Astrophysical Journal Supplement Series},
	month = {jan},
	number = {1},
	pages = {31},
	title = {Simons Observatory: Predeployment Performance of a Large Aperture Telescope Optics Tube in the 90 and 150 GHz Spectral Bands},
	volume = {276},
	year = {2025}}

@phdthesis{Huber:2025elf,
    author = "Huber, Zachary",
    title = "{Constraining Time-Dependent Parity Violation with ACT and Developing Next-Generation Microwave Observatories}",
    doi = "10.7298/72am-9p63",
    school = "Cornell U.",
    year = "2025"
}

@MISC{ALMA_handbook,
       author = {{Remijan}, A. and {Biggs}, A. and {Cortes}, P.~A. and {Dent}, B. and {Di Franceso}, J. and {Fomalont}, E. and {Hales}, A. and {Kameno}, S. and {Mason}, B. and {Philips}, N. and {Saini}, K. and {Vila Vilaro}, B. and {Villard}, E.},
        title = "{ALMA Technical Handbook,ALMA Doc. 7.3, ver. 1.1}",
 howpublished = {2019, ALMA Technical Handbook,ALMA Doc. 7.3, ver. 1.1ISBN 978-3-923524-66-2},
         year = 2019,
        month = jun,
          doi = {10.5281/zenodo.4511522},
       adsurl = {https://ui.adsabs.harvard.edu/abs/2019athb.rept.....R}}

@article{Xu_metamaterial,
	author = {Zhilei Xu and Grace E. Chesmore and Shunsuke Adachi and Aamir M. Ali and Andrew Bazarko and Gabriele Coppi and Mark Devlin and Tom Devlin and Simon R. Dicker and Patricio A. Gallardo and Joseph E. Golec and Jon E. Gudmundsson and Kathleen Harrington and Makoto Hattori and Anna Kofman and Kenji Kiuchi and Akito Kusaka and Michele Limon and Frederick Matsuda and Jeff McMahon and Federico Nati and Michael D. Niemack and Aritoki Suzuki and Grant P. Teply and Robert J. Thornton and Edward J. Wollack and Mario Zannoni and Ningfeng Zhu},
	journal = {Appl. Opt.},
	month = {Feb},
	number = {4},
	pages = {864--874},
	title = {The Simons Observatory: metamaterial microwave absorber and its cryogenic applications},
	volume = {60},
	year = {2021}}

@INPROCEEDINGS{hamdi_amps,
  author={Mani, Hamdi and Mauskopf, Philip},
  booktitle={2014 11th International Workshop on Low Temperature Electronics (WOLTE)}, 
  title={A single-stage cryogenic LNA with low power consumption using a commercial SiGe HBT}, 
  year={2014},
  volume={},
  number={},
  pages={17-20},
  doi={10.1109/WOLTE.2014.6881015}}

@ARTICLE{keller_stripline,
  author={Keller, Ben and Freundt, Rodrigo and Burgoyne, James R. and Chapman, Scott and Choi, Steve and Duell, Cody J. and Groppi, Christopher and Humphreys, Caleb and Lin, Lawrence T. and Middleton, Alicia and Niemack, Michael D. and Patel, Darshan and Vavagiakis, Eve and Walker, Samantha and Wang, Yuhan and Xie, Ruixuan},
  journal={IEEE Transactions on Applied Superconductivity}, 
  title={CCAT: Flexible Stripline Circuits for Large-Format Kinetic Inductance Detector (KID) Array Readout}, 
  year={2026},
  volume={36},
  number={6},
  pages={1-6},
  doi={10.1109/TASC.2026.3669460}}

@inproceedings{sinclair_rfsoc,
	author = {Adrian K. Sinclair and Ryan C. Stephenson and Cody A. Roberson and Eric L. Weeks and James Burgoyne and Anthony I. Huber and Philip M. Mauskopf and Scott C. Chapman and Jason E. Austermann and Steve K. Choi and Cody J. Duell and Michel Fich and Christopher E. Groppi and Zachary Huber and Michael D. Niemack and Thomas Nikola and Kayla M. Rossi and Adhitya Sriram and Gordon J. Stacey and Erik Szakiel and Joel Tsuchitori and Eve M. Vavagiakis and Jordan D. Wheeler},
	booktitle = {Millimeter, Submillimeter, and Far-Infrared Detectors and Instrumentation for Astronomy XI},
	editor = {Jonas Zmuidzinas and Jian-Rong Gao},
	organization = {International Society for Optics and Photonics},
	pages = {121900W},
	publisher = {SPIE},
	title = {{CCAT-prime: RFSoC based readout for frequency multiplexed kinetic inductance detectors}},
	volume = {12190},
	year = {2022}}

@inproceedings{gascard_thermal_model,
	author = {Thomas J. L. J. Gascard and Yi Wang and Jon E. Gudmundsson and Eve M. Vavagiakis and Cody J. Duell and Zachary B. Huber and Lawrence T. Lin and Michael D. Niemack and Rodrigo G. Freundt},
	booktitle = {Millimeter, Submillimeter, and Far-Infrared Detectors and Instrumentation for Astronomy XII},
	editor = {Jonas Zmuidzinas and Jian-Rong Gao},
	organization = {International Society for Optics and Photonics},
	pages = {131022O},
	publisher = {SPIE},
	title = {{Thermal and mechanical study of a parametrised cryostat model for optical characterisation of upcoming CMB experiments}},
	volume = {13102},
	year = {2024}}

@ARTICLE{middleton_LEDmap,
  author={Middleton, Alicia and Choi, Steve K. and Walker, Samantha and Austermann, Jason and Burgoyne, James R. and Butler, Victoria and Chapman, Scott C. and Crites, Abigail T. and Duell, Cody J. and Freundt, Rodrigo G. and Huber, Anthony I. and Huber, Zachary B. and Hubmayr, Johannes and Keller, Ben and Lin, Lawrence T. and Niemack, Michael D. and Patel, Darshan and Sinclair, Adrian K. and Smith, Ema and Vaskuri, Anna and Vavagiakis, Eve M. and Vissers, Michael and Wang, Yuhan and Wheeler, Jordan},
  journal={IEEE Transactions on Applied Superconductivity}, 
  title={CCAT: LED Mapping and Characterization of the 280 GHz TiN KID Array}, 
  year={2025},
  volume={35},
  number={5},
  pages={1-4},
  doi={10.1109/TASC.2024.3517564}}

@article{pixie_design,
	author = {Kogut, Alan and Aghanim, Nabila and Chluba, Jens and Chuss, David T. and Delabrouille, Jacques and Dvorkin, Cora and Fixsen, Dale and Ghosh, Shamik and Hensley, Brandon S. and Hill, J. Colin and Maffei, Bruno and Pullen, Anthony R. and Rotti, Aditya and Sabyr, Alina and Switzer, Eric R. and Thiele, Leander and Wollack, Edward J. and Zelko, Ioana},
	journal = {Journal of Cosmology and Astroparticle Physics},
	month = {apr},
	number = {04},
	pages = {020},
	title = {The Primordial Inflation Explorer (PIXIE): mission design and science goals},
	volume = {2025},
	year = {2025}}

@inproceedings{duell_280ghz,
	author = {Cody J. Duell and Eve M. Vavagiakis and Jason Austermann and Scott C. Chapman and Steve K. Choi and Nicholas F. Cothard and Brad Dober and Patricio Gallardo and Jiansong Gao and Christopher Groppi and Terry L. Herter and Gordon J. Stacey and Zachary Huber and Johannes Hubmayr and Doug Johnstone and Yaqiong Li and Philip Mauskopf and Jeff McMahon and Michael D. Niemack and Thomas Nikola and Kayla Rossi and Sara Simon and Adrian K. Sinclair and Michael Vissers and Jordan Wheeler and Bugao Zou},
	booktitle = {Millimeter, Submillimeter, and Far-Infrared Detectors and Instrumentation for Astronomy X},
	editor = {Jonas Zmuidzinas and Jian-Rong Gao},
	organization = {International Society for Optics and Photonics},
	pages = {114531F},
	publisher = {SPIE},
	title = {{CCAT-prime: Designs and status of the first light 280 GHz MKID array and mod-cam receiver}},
	volume = {11453},
	year = {2020}}

@article{vaskuri_280ghz,
	author = {Anna K. Vaskuri and Jordan D. Wheeler and Jason E. Austermann and Michael R. Vissers and James A. Beall and James Burgoyne and Victoria Butler and Scott Chapman and Steve K. Choi and Abigail Crites and Cody J. Duell and Rodrigo Freundt and Anthony Huber and Zachary B. Huber and Johannes Hubmayr and Jozsef Imrek and Ben Keller and Lawrence Lin and Alicia Middleton and Michael D. Niemack and Thomas Nikola and Douglas Scott and Adrian Sinclair and Ema Smith and Gordon Stacey and Joel Ullom and Jeffrey van Lanen and Eve M. Vavagiakis and Samantha Walker and Bugao Zou},
	journal = {Journal of Astronomical Telescopes, Instruments, and Systems},
	number = {2},
	pages = {026005},
	title = {{280-GHz aluminum MKID arrays for the Fred Young Submillimeter Telescope}},
	volume = {11},
	year = {2025}}

@article{blast_dets,
	author = {Dober, B. and Austermann, J. A. and Beall, J. A. and Becker, D. and Che, G. and Cho, H. M. and Devlin, M. and Duff, S. M. and Galitzki, N. and Gao, J. and Groppi, C. and Hilton, G. C. and Hubmayr, J. and Irwin, K. D. and McKenney, C. M. and Li, D. and Lourie, N. and Mauskopf, P. and Vissers, M. R. and Wang, Y.},
	journal = {Journal of Low Temperature Physics},
	number = {1},
	pages = {173--179},
	title = {Optical Demonstration of THz, Dual-Polarization Sensitive Microwave Kinetic Inductance Detectors},
	volume = {184},
	year = {2016}}

@article{toltec_dets,
	author = {Austermann, J. E. and Beall, J. A. and Bryan, S. A. and Dober, B. and Gao, J. and Hilton, G. and Hubmayr, J. and Mauskopf, P. and McKenney, C. M. and Simon, S. M. and Ullom, J. N. and Vissers, M. R. and Wilson, G. W.},
	journal = {Journal of Low Temperature Physics},
	number = {3},
	pages = {120--127},
	title = {Millimeter-Wave Polarimeters Using Kinetic Inductance Detectors for TolTEC and Beyond},
	volume = {193},
	year = {2018}}

@ARTICLE{austermann_feeds,
  author={Austermann, Jason and Beall, James and Burgoyne, James R. and Chapman, Scott and Choi, Steve and Duell, Cody J. and Huber, Anthony I. and Hubmayr, Johannes and Koc, Matthew A. and Niemack, Michael D. and Ullom, Joel N. and van Lanen, Jeffrey and Vaskuri, Anna and Vissers, Michael and Wheeler, Jordan},
  journal={IEEE Transactions on Applied Superconductivity}, 
  title={CCAT: Silicon-Platelet Feedhorns for Submillimeter Wavelengths}, 
  year={2026},
  volume={36},
  number={6},
  pages={1500708-1500708},
  doi={10.1109/TASC.2026.3689810}}

@ARTICLE{pwv_study,
       author = {{Cort{\'e}s}, F. and {Cort{\'e}s}, K. and {Reeves}, R. and {Bustos}, R. and {Radford}, S.},
        title = "{Twenty years of precipitable water vapor measurements in the Chajnantor area}",
      journal = {AAP},
         year = 2020,
        month = aug,
       volume = {640},
          eid = {A126},
        pages = {A126},
          doi = {10.1051/0004-6361/202037784},
archivePrefix = {arXiv},
       eprint = {2007.04262},
 primaryClass = {astro-ph.IM},
       adsurl = {https://ui.adsabs.harvard.edu/abs/2020A&A...640A.126C}
}

@article{liu_ledmap,
    author = {Liu, X. and Guo, W. and Wang, Y. and Wei, L. F. and Mckenney, C. M. and Dober, B. and Billings, T. and Hubmayr, J. and Ferreira, L. S. and Vissers, M. R. and Gao, J.},
    title = {Cryogenic LED pixel-to-frequency mapper for kinetic inductance detector arrays},
    journal = {Journal of Applied Physics},
    volume = {122},
    number = {3},
    pages = {034502},
    year = {2017},
    month = {07},
    issn = {0021-8979},
    doi = {10.1063/1.4994170},
    url = {https://doi.org/10.1063/1.4994170},
    eprint = {https://pubs.aip.org/aip/jap/article-pdf/doi/10.1063/1.4994170/15196695/034502_1_online.pdf},
}

@unpublished{vavagiakis_pcam_inprep,
  title = {{{CCAT}}: The Prime-Cam Instrument for the Fred Young Submillimeter Telescope – Overview and Status},
  author = {Vavagiakis, E. and Wang, Y. and Lin, L.T. and others},
  year = 2026,
  note = {Manuscript in preparation},
  langid = {english}
}

@unpublished{tilak_450_inprep,
  title = {{{CCAT}}: Optical Design of the 410 GHz Prime-Cam Module},
  author = {Tilak M. Patel and Tanner Buck and Anthony Huber and Eve M. Vavagiakis and James Burgoyne and Scott Chapman and Ben Keller and Michael D. Niemack and {CCAT Collaboration}},
  year = 2026,
  note = {Manuscript in preparation},
  langid = {english}
}

@inproceedings{rodrigo_eor,
	author = {Rodrigo Freundt and Yaqiong Li and Doug Henke and Jason Austermann and James R. Burgoyne and Scott Chapman and Steve K. Choi and Cody J. Duell and Zach Huber and Michael Niemack and Thomas Nikola and Lawrence Lin and Dominik A. Riechers and Gordon Stacey and Anna K. Vaskuri and Eve M. Vavagiakis and Jordan Wheeler and Bugao Zou},
	booktitle = {Millimeter, Submillimeter, and Far-Infrared Detectors and Instrumentation for Astronomy XII},
	editor = {Jonas Zmuidzinas and Jian-Rong Gao},
	organization = {International Society for Optics and Photonics},
	pages = {131020U},
	publisher = {SPIE},
	title = {{CCAT: a status update on the EoR-Spec instrument module for Prime-Cam}},
	volume = {13102},
	year = {2024}}

@inproceedings{chapman_850,
	author = {Scott C. Chapman and Anthony I. Huber and Adrian K. Sinclair and Jordan D. Wheeler and Jason E. Austermann and James Beall and James Burgoyne and Steve K. Choi and Abigail Crites and Cody J. Duell and Jesslyn Devina and Jiansong Gao and Mike Fich and Doug Henke and Terry Herter and Doug Johnstone and Lewis B. G. Knee and Michael D. Niemack and Kayla M. Rossi and Gordon Stacey and Joel Tsuchitori and Joel Ullom and Jeff Van Lanen and Eve M. Vavagiakis and Michael Vissers},
	booktitle = {Millimeter, Submillimeter, and Far-Infrared Detectors and Instrumentation for Astronomy XI},
	editor = {Jonas Zmuidzinas and Jian-Rong Gao},
	organization = {International Society for Optics and Photonics},
	pages = {1219005},
	publisher = {SPIE},
	title = {{CCAT-prime: the 850 GHz camera for prime-cam on FYST}},
	volume = {12190},
	year = {2022}}

@article{pan_pixie,
	author = {Zhaodi Pan and Mira Liu and Ritoban Basu Thakur and Bradford A. Benson and Dale J. Fixsen and Hazal Goksu and Eleanor Rath and Stephan S. Meyer},
	journal = {Appl. Opt.},
	month = {Aug},
	number = {23},
	pages = {6257--6267},
	title = {Compact millimeter-wavelength Fourier-transform spectrometer},
	volume = {58},
	year = {2019}}

@misc{amcode_model,
  author       = {Paine, Scott},
  title        = {The am atmospheric model},
  month        = sep,
  year         = 2024,
  publisher    = {Zenodo},
  version      = {14.0},
  doi          = {10.5281/zenodo.13748391},
  url          = {https://doi.org/10.5281/zenodo.13748391},
}

@article{Noroozian_TLS,
    author = {Noroozian, Omid and Gao, Jiansong and Zmuidzinas, Jonas and LeDuc, Henry G. and Mazin, Benjamin A.},
    title = {Two‐level system noise reduction for Microwave Kinetic Inductance Detectors},
    journal = {AIP Conference Proceedings},
    volume = {1185},
    number = {1},
    pages = {148-151},
    year = {2009},
    month = {12},
    issn = {0094-243X},
    doi = {10.1063/1.3292302},
    url = {https://doi.org/10.1063/1.3292302},
    eprint = {https://pubs.aip.org/aip/acp/article-pdf/1185/1/148/12248565/148_1_online.pdf},
}
\bibliographystyle{spiebib} 

\end{document}